\documentclass[aps,showpacs,pra,twocolumn,longbibliography]{revtex4-1}

\usepackage[normalem]{ulem}
\usepackage{exscale}
\usepackage{graphicx}
\usepackage{amsmath}
\usepackage{latexsym}
\usepackage{amsfonts}
\usepackage{amssymb}
\usepackage{times}
\usepackage[T1]{fontenc}
\usepackage{amsthm}
\usepackage{enumerate}
\usepackage{bbold}
\usepackage{color}
\usepackage{nicefrac}
\usepackage[colorlinks=true,citecolor=blue,urlcolor=blue]{hyperref}
\usepackage{mathtools}

\newcommand{\ba}{\begin{eqnarray}}
\newcommand{\be}{\begin{equation}}
\newcommand{\ee}{\end{equation}}
\newcommand{\beq}{\begin{equation}}
\newcommand{\eeq}{  \end{equation}}
\newcommand{\bea}{\begin{eqnarray}}
\newcommand{\eea}{  \end{eqnarray}}

\newcommand{\ea}{\end{eqnarray}}
\newcommand{\ban}{\begin{eqnarray*}}
\newcommand{\ean}{\end{eqnarray*}}

\newcommand{\tr}{\operatorname{tr}}

\newcommand{\ket}[1]{\left|#1\right\rangle}

\newcommand{\ie}{{\it{i.e.}~}}

\begin{document}

\title{Efficient Device-Independent Entanglement Detection for Multipartite Systems}

%%%%%%%%%%%%%%%%%%%%%%%%%%%%%%%%%%%%%%%%%%%%%%%%%%%%%%%%%%%%%%%%%%

\author{F. Baccari$^1$}
\email{flavio.baccari@icfo.eu}
%\author{Mari Cobra$^1$}
\author{D. Cavalcanti$^1$}
\author{P. Wittek$^{1,2}$}
\author{A. Ac\'{i}n$^{1,3}$ }
\affiliation{$^1$ICFO - Institut de Ciences Fotoniques, The Barcelona Institute of Science and Technology, 08860 Castelldefels, Spain\\
$^2$ University of Bor\r{a}s, 50190 Bor\r{a}s, Sweden \\
$^3$ICREA - Instituci\'{o} Catalana de Recerca i Estudis Avan\c{c}ats, Lluis Campanys 23, 08010 Barcelona, Spain}

%%%%%%%%%%%%%%%%%%%%%%%%%%%%%%%%%%%%%%%%%%%%%%%%%%%%%%%%%%%%%%%%%%
\begin{abstract}
Entanglement is one of the most studied properties of quantum
mechanics for its application in quantum information
protocols. Nevertheless, detecting the presence of
entanglement in large multipartite sates continues to be a great
challenge both from the theoretical and the experimental point of
view. Most of the known methods either have computational costs
that scale inefficiently with the number of particles or require
more information on the state than what is attainable in everyday
experiments. We introduce a new technique for entanglement
detection that provides several important advantages in
these respects. First, it scales efficiently with the number of
particles, thus allowing for application to systems composed by up
to few tens of particles. Second, it needs only the knowledge of
a subset of all possible measurements on the state,
therefore being apt for experimental implementation. Moreover,
since it is based on the detection of nonlocality, our method is
device independent. We report several examples of its
implementation for well-known multipartite states, showing that
the introduced technique has a promising range of applications.
\end{abstract}

%%%%%%%%%%%%%%%%%%%%%%%%%%%%%%%%%%%%%%%%%%%%%%%%%%%%%%%%%%%%%%%%%

\maketitle

%%%%%%%%%%%%%%%%%%%%%%%%%%%%%%%%%%%%%%%%%%%%%%%%%%%%%%%%%%%%%%%
\section{Introduction}

Entanglement is the key ingredient for several protocols
in quantum information theory, such as quantum
teleportation~\cite{bennett1993teleporting}, quantum key
distribution~\cite{ekert1991quantum}, measurement-based quantum
computation~\cite{briegel2009measurement} and quantum metrology
schemes~\cite{giovannetti2004quantum}. Therefore, developing
techniques to detect the presence of entanglement in quantum
states is crucial and in the past years several methods have been
introduced. 

The most general way to detect entanglement in a given system
consists of reconstructing its quantum state using tomography and
then applying any entanglement criterion to the
resulting state ~\cite{horodecki2009entanglement}. This, however,
is costly both from an experimental and a theoretical perspective.
First, determining the state of large quantum systems is
impractical in experiments, given that quantum tomography implies
measuring a number of observables that increases
exponentially with the number of systems, e.g., $3^N$ observables
even in the simplest case of $N$
qubits~\cite{haffner2005scalable}. Second, determining whether an
arbitrary state is entangled is known to be a hard problem -- to
the best of our knowledge, the computational resources of the most
efficient known algorithm scale exponentially with $N$
~\cite{brandao2012detection}. Because of these problems,
it is very desirable to develop entanglement detection techniques with more accessible experimental and computational requirements.

One possible approach is to make use of entanglement witnesses.
These are criteria for detecting entanglement that require
measuring only some expectation values of local
observables~\cite{bourennane2004experimental}. 
In particular, attempts have been made to derive witnesses that adapt to the limited amount of information that is usually available in a typical experiment. For instance, one can consider witnesses
involving only two-body correlators~\cite{toth2006detection} or a
few global
measurements~\cite{lucke2014detecting,vitagliano2014spin}.
Nonetheless, entanglement witnesses constitute a method that lacks
generality, given that the known methods are generally tailored
to detect very specific states.
There are techniques capable of deriving a witness for any generic entangled state, which can also be constrained to the available set of data~\cite{jungnitsch2011taming}, or adapted to require the minimal amount of measurements on the system \cite{knips2014multipartite}. However, they always involve an optimization procedure that runs on an exponentially increasing number of parameters.  
A method to detect metrologically useful (hence entangled) states 
based on a couple of measurements has recently been proposed \citep{apellaniz2015optimal}. However, these states represent only a subset of all entangled states.

A qualitatively different approach to entanglement detection is based on Bell
nonlocality~\citep{bell1964on}. Indeed, the presence of nonlocality
provides a certificate of the entanglement in the state. Moreover,
it has the advantage that it can be assessed in a
device-independent manner, i.e. without making any assumption on
the actual experimental implementation~\cite{brunner2014bell}. 
The easiest way to detect nonlocality is by means of the violation of a Bell inequality. However, in analogy with the entanglement case, each inequality is usually violated by a very specific class of states.
In the general case, verifying whether a set of observed correlations
is nonlocal can be done via linear
programing~\citep{zukowski1999strengthening}. Nonetheless, the number of variables involved again grows exponentially with the number of particles, e.g. as $4^N$ already for the simplest
scenarios where only two dichotomic measurements per party are
applied~\cite{brunner2014bell}.

To summarize, the methods to detect entanglement known so far are either not general or too costly, from a computational and/or experimental viewpoint, to be applied to large systems. 

Here we present a novel technique for device-independent entanglement detection that is efficient both experimentally and
computationally. On the one hand, it requires the knowledge of
a subset of all possible measurements, most of them consisting of 
few-body correlation functions, which makes it suitable for practical
implementations. On the other hand, it can be applied to any set
of observed correlations and can be implemented by semidefinite
programing involving a number of variables that grows
polynomially with $N$. 

Of course, all these nice properties
become possible only because our method for entanglement detection is
a relaxation of the initial hard problem. However, and despite being a relaxation, we demonstrate the power of our approach by showing 
how it can be successfully applied to several physically relevant examples for systems of up to $29$ qubits. These examples demonstrate that our approach opens a promising avenue for entanglement detection of large many-body quantum systems.

This article is organized as follows: in Section~\ref{sec:notation}
we introduce the basic notation and definitions, while in Section \ref{sec:method} we present the idea of the method together with the application to a simple scenario.
Section~\ref{sec:geometrical} is devoted to the presentation of
its geometrical interpretation together with the comparison to the
other techniques. In Section~\ref{sec:results} and~\ref{sec:bell} we list some
examples of application of the method to relevant classes of
states. Lastly, Section~\ref{sec:conclusions} contains conclusions
and some future perspectives.

%%%%%%%%%%%%%%%%%%%%%%%%%%%%%%%%%%%%%%%%%%%%%%%%%%%%%%%%%%%%%%%%%%
\section{Notation and Definitions}
\label{sec:notation}

We consider an entanglement detection scenario in which $N$
observers, denoted by $A_1,{\ldots},A_N$, share an $N$-partite
quantum state $\rho_N$. Each $A_i$ performs $m$ possible
measurements, each having $d$ outcomes. We represent the
measurement of party $i$ by ${M}_{x_i}^{a_i}$, where $x_i \in
\lbrace0,{\ldots},m-1 \rbrace$ denotes the measurement choice and
$a_i \in \lbrace 0,{\ldots},d-1 \rbrace$ are the corresponding
outcomes.

By repeating the experiment sufficiently many times, the observers
can estimate the conditional probabilities
\begin{equation}
p(a_1,{\ldots},a_N | x_1,{\ldots},x_N) = \tr(M_{x_1}^{a_1} \otimes {\ldots} \otimes M_{x_N}^{a_N} \rho_N),
\label{quantum}
\end{equation}
of getting the different outcomes depending on the measurements
they have performed. The conditional probability
distribution $p(a_1,{\ldots},a_N | x_1,{\ldots},x_N)$ describes the
correlations observed among the observers when applying the local
measurements ${M}_{x_i}^{a_i}$ on the state $\rho_N$.

One can also define the marginal distributions
\begin{equation}
p(a_{i_1},{\ldots},a_{i_k} | x_{i_1},{\ldots}x_{i_k}) = \tr(M_{x_{i_1}}^{a_{i_1}} \otimes {\ldots} \otimes M_{x_{i_k}}^{a_{i_k}} \rho_{i_1,{\ldots},i_k})
\label{marginals}
\end{equation}
where $0 \leq i_1 < {\ldots} < i_k < N$, $1 \leq k \leq N$ and
$\rho_{i_1,{\ldots},i_k}$ is the reduced state of $\rho_N$
corresponding to the considered subset of parties. Marginals can
equivalently be obtained from the full distribution
(\ref{quantum}) by summing over the remaining outcomes.

Since in what follows we mostly consider scenarios involving two-output measurements (resulting from projective measurements performed on qubits), it is convenient to introduce the concept of correlators 

\begin{widetext}
\begin{equation}
\langle M_{x_{i_1}}^{(i_1)}{\ldots} M_{x_{i_k}}^{(i_k)}   \rangle = \sum_{a_{i_1},{\ldots},a_{i_k}} (-1)^{\sum_{l =1}^k a_{i_l}} p(a_{i_1}{\ldots},a_{i_k}| x_{i_1},{\ldots},x_{i_k})
\label{corr}
\end{equation}
\end{widetext}
where $0 \leq i_1 < {\ldots} < i_k < N$, $x_{i_j} \in \lbrace 0,m-1 \rbrace$ and $1 \leq k \leq N$. The value of $k$ represents the order of the correlators: for instance, expectation values $\langle M_{x_{i_1}}^{(i_1)} M_{x_{i_2}}^{(i_2)} \rangle$ are of order two. Correlators of order $N$ are often referred to as full-body correlators.
In scenarios involving only dicothomic measurements, correlators encode all the information in the observed distribution~\eqref{quantum}.
When working with correlators, it is also useful to introduce the measurement operators in the expectation value form, namely by using the notation $M^{(i)}_{x_i} = M^{1}_{x_i} - M^{0}_{x_i}$. With this definition, it is easy to see that $\langle M_{x_{i_1}}^{(i_1)}{\ldots} M_{x_{i_k}}^{(i_k)} \rangle = \tr(M_{x_{i_1}}^{(i_1)} \otimes {\ldots} \otimes M_{x_{i_k}}^{(i_k)} \rho_{i_1,{\ldots},i_k})$.

Now that the main concepts have been introduced, we proceed with outlining the proposed entanglement detection method.

%%%%%%%%%%%%%%%%%%%%%%%%%%%%%%%%%%%%%%%%%%%%%%%%%%%%%%%%%%%%%%%%%%%%%

\section{Method}
\label{sec:method}

Our method is based on the following reasoning (discussed in detail below):
\begin{enumerate}
\item If a quantum state $\rho_N$ is separable, local measurements performed on it produce local correlations (i.e. correlations admitting a local model).

\item Any local correlations can be realized by performing commuting local measurements on a quantum state.

\item Correlations produced by commuting local measurements define a positive moment matrix with constraints associated to the commutation of all the measurements.

\item Our method consists in checking if the observed correlations are consistent with such a positive moment matrix. In the negative case the state $\rho_N$ producing the correlations is proven to be entangled.
\end{enumerate}

Let us now explain all these points in detail.

First, given a separable quantum state, \ie
$\rho_N=\sum_\lambda p_\lambda \bigotimes_i \rho^{A_i}_\lambda$,
any set of conditional probability distributions obtained
after performing local measurements on it admits a decomposition
of the following form \ba
p(a_1,{\ldots},a_N | x_1,{\ldots}, x_N) &=& \sum_\lambda p_\lambda\tr(\bigotimes_i M_{x_i}^{a_i} \bigotimes_i \rho^{A_i}_\lambda)\nonumber\\
&=& \sum_\lambda p_\lambda \prod_{i = 1}^N p(a_i | x_i , \lambda),
\label{local} \label{localmodel} \ea where $p(a_i | x_i , \lambda)=\tr(M_{x_i}^{a_i}
\rho^{A_i}_\lambda)$. In the context of Bell nonlocality,
distributions that can be written in this form are called
local~\cite{brunner2014bell}. Local correlations do not
violate any Bell inequality. 
Conversely, if a given distribution cannot be described by a local model like (\ref{localmodel}), it is said to be nonlocal. We notice that, by the reasoning presented above, whenever the set of observed distributions
(\ref{quantum}) is nonlocal, we can conclude that the shared state
is entangled.
Moreover, since nonlocality is a property that can be assessed at the level of the probability distribution, it can be seen as a device-independent way of detecting entanglement. For the sake of brevity, throughout the rest of the paper we therefore refer to our method as a nonlocality detection one.

The second ingredient is that any local set of probability
distributions has a quantum realization in terms of local
\emph{commuting measurements} applied to a quantum
state~\cite{fine1982hidden}. In order to see it more explicitly we
first realize that any decomposition of the form (\ref{local}) can
be rewritten as
\begin{equation}
p(a_1,{\ldots},a_N | x_1,{\ldots}, x_N) = \sum_{\lambda} q_{\lambda} \prod_{i = 1}^{N} D(a_i | x_i , \lambda)
\label{deter}
\end{equation}
where $D(a_i | x_i , \lambda)$ are deterministic functions that give a fixed outcome $a$ for each measurement, \ie $D(a_i|x_i,\lambda) = \delta_{a_i,\lambda(x_i)}$, such that $a_i = \lambda(x_i)$, with $\lambda(\cdot)$ a function from $\{0,\ldots,m-1\}$ to $\{0,\ldots, d-1\}$~\cite{brunner2014bell}. It is easy to see that any such decomposition can be reproduced by choosing the multipartite state $\rho_{N} = \sum_{\lambda} q_{\lambda} | \lambda \rangle \langle \lambda |^{\otimes N}$ and measurement operators of the form $M_{x_i}^{a_i} = \sum_{\lambda{'}} D(a_i | x_i , \lambda{'} )  | \lambda{'} \rangle \langle \lambda{'} |$. In particular, $[ M_{x_i}^{(i)} , M_{x'_i}^{(i)} ] = 0 ~\forall ~ i, x_i,$ and $x'_i$.

The last step consists in using a modified version of the
Navascues-Pironio-Acin (NPA)
hierarchy~\cite{navascues2007bounding,navascues2008convergent}
that takes into account the commutativity of the local measurements to
test if the observed probability distribution is local (a similar
idea was introduced in the context of quantum
steering~\cite{kogias2015hierarchy} -- see
also~\cite{cavalcanti2016quantum}). The NPA hierarchy
consists of a sequence of tests aimed at certifying if a given set
of probability distributions has a quantum realization
\eqref{quantum}. In NPA one imposes the commutativity of the
measurements between the distant parties. Now, we will impose the
extra constraints that the local measurements on each party also
commute. The resulting semidefinite program (SDP) hierarchy is nothing but an application
in this context of the more general method for polynomial
optimization over noncommuting variables introduced
in~\cite{PNA}, see also~\cite{moments}. As noticed there, by
imposing commutativity of all the variables this general hierarchy
reduces to the well-known Lassere hierarchy, namely the
relaxation for polynomial optimization of commuting
variables~\cite{lasserre2001global}. An application of this
relaxation technique to describe local correlations was also
proposed in~\cite{steeg2011sequence}. However, to the best of our
knowledge, no systematic analysis of its application to
multipartite scenarios has been considered thus far.

\subsection*{Details and convergence of the hierarchy}

It is convenient for what follows to recall the main
ingredients of the NPA
hierarchy~\cite{navascues2007bounding,navascues2008convergent},
which, as said, was designed to characterize probability
distributions with a quantum realization~\eqref{quantum}.
Consider a set $\mathcal{O}$, composed by some products of the
measurements operators  $\lbrace M_{x_i}^{a_i} \rbrace$ or linear
combinations of them. By indexing the elements in the set as
$\mathcal{O}_i$ with $i = 1,{\ldots}k$, we introduce the so-called
moment matrix $\Gamma$ as the $k \times k$ matrix whose entries
are defined by $ \Gamma_{ij} = tr (\rho_N \mathcal{O}^{\dagger}_i
\mathcal{O}_j)$. For any choice of measurements and state, it can
be shown that $\Gamma$ satisfies the following properties: i) it
is positive semidefinite, ii) its entries satisfy a series of
linear constraints associated to the commutation relations
among measurement operators by different parties and the fact that
they correspond to projectors, iii) some of its entries can be computed
from the observed probability distribution (\ref{quantum}), 
iv) some of its entries correspond to
unobservable numbers (e.g. when $O_i$ and $O_j$ involve
noncommuting observables). 

Based on these facts one can define a
hierarchy of tests to check whether a given set of correlations
has a quantum realization. One first defines the sets
$\mathcal{O}_{\nu}$ composed of products of at most $\nu$ of the
measurement operators, and creates the corresponding $\Gamma$
matrix using the set of correlations and leaving the unassigned entries
as variables. Then one seeks for values for these variables that
could make the $\Gamma$ positive. This problem constitutes a
SDP, for which some efficient solving
algorithms are known~\cite{cleve2004consequences}. If no such
values are found this means that the set of correlations used does
not have a quantum realization. By increasing the value of $\nu$,
one gets a sequence of stricter and stricter ways of testing the
belonging of a distribution to the quantum set.

We can now use the same idea to define a hierarchy of conditions
to test whether a given set of correlations has a quantum
realization with commuting measurements. To do so we simply
impose additional linear constraints on the entries of the moment
matrix resulting from assuming that the local measurements
also commute (for a more detailed discussion, see
Appendix~\ref{app:details}). Thus, given a set of observed
probability distributions one can use them to build a
NPA-type matrix with the additional linear constraints
associated to the local commutation relations, and run a
SDP to check its positivity to certify if the considered set of
correlations can not be obtained by measuring a separable state.

Interestingly, the convergence of this hierarchy follows from the results in \cite{navascues2008convergent,PNA}.
Roughly speaking, one can say that since the NPA hierarchy is proven to converge to the set of quantum correlations, our method provides a hierarchy that converges to the set of quantum correlations with commuting measurements, which we have shown to be equivalent to the set of local correlations \footnote{A more formal way to see it is to notice that we are restricting the projective algebra of NPA by the commutation condition. Then, given that the original algebra already meets the Archimedean condition, the convergence holds for the commuting case as well.}.
Therefore, any nonlocal correlation would fail the SDP test at a finite step of the sequence given by the $\mathcal{O}_\nu$.

Moreover, the commutativity of all the measurements implies that the total number of variables that can be involved in the SDP test is finite. The reason is that the longest nontrivial product of the operators that can appear in the moment matrix consists of the products of all the different $M_{x_i}^{a_i}$. Hence, the number of variables in the moment matrix stops growing after the first step at which this product appears. This also implies that the convergence of the hierarchy is met at a finite step as well, namely coinciding to the level $\nu{'}$ at which the longest nontrivial products appear in the list of operators $\mathcal{O}_{\nu{'}}$. Indeed, it is easy to see that for $\mu > \nu{'}$, there cannot appear new operators in the generating set, i.e. $\mathcal{O}_{\mu} = \mathcal{O}_{\nu{'}}$.
Of course, the aforementioned levels depend on the numbers $(N,m,d)$ defining the scenario and it is, in general, high. Indeed, according to the Collins-Gisin representation \cite{collins2004arelevant}, one has $N m (d-1)$ independent measurement operators $M_{x_i}^{a_i}$. Therefore, the product of all of them would first appear in the moment matrix at level $\lceil \frac{N m (d-1)}{2} \rceil $. Consequently, convergence is assured at level $\nu' = N m (d-1)$. 

To conclude, we stress that, depending on the level of the hierarchy, one might not need
knowledge of the full probability distribution. Indeed, by looking
at (\ref{marginals}), it is evident that to define a marginal
distribution involving $k$ parties, one requires the product of $k$
measurements $M_{x_i}^{a_i}$. Now, given that the operators of the
set $\mathcal{O}_{\nu}$ contain products of at most $\nu$
measurement operators, the terms in the moment matrix at level
$\nu$ can only coincide with the marginals of the observed
distribution of up to $k = 2 \nu$ parties. Therefore, in the
multipartite setting, fixing the level of the hierarchy is also a
way to limit the order of the marginals that can be assigned in
the moment matrix.

\subsection*{Simple example}

After presenting the general idea of the method, it is
convenient to conclude the section by illustrating it with a concrete example.
In what follows, we present the explicit form of the moment matrix for the bipartite case, two dicothomic measurements per party and level $\nu = 2$ of
the hierarchy. For the sake of simplicity, we rename the
expectation value operators for the two parties as $\mathcal{A}_x$ and
$\mathcal{B}_y$, with $x,y = 0,1$. In this scenario, the set of
operators reads as $\mathcal{O}_{2} = \lbrace \mathbb{1},
\mathcal{A}_0, \mathcal{A}_1 ,\mathcal{B}_0 , \mathcal{B}_1 ,
\mathcal{A}_0 \mathcal{A}_1,  \mathcal{A}_0 \mathcal{B}_0 ,
\mathcal{A}_0 \mathcal{B}_1,  \mathcal{A}_1 \mathcal{B}_0 ,
\mathcal{A}_1 \mathcal{B}_1,  \mathcal{B}_0 \mathcal{B}_1
\rbrace$. The corresponding moment matrix is

\begin{widetext}

\begin{equation}
\Gamma = \left(
\begin{array}{ccccccccccc}

1 & \textcolor{red}{\langle \mathcal{A}_0 \rangle} &   \langle \mathcal{A}_1 \rangle &   \textcolor{blue}{\langle \mathcal{B}_0 \rangle} &   \langle \mathcal{B}_1 \rangle & \color[rgb]{0.870364,0.870097,0.000000}{v_1} &  \color[rgb]{0.250980,0.501961,0.000000}{\langle \mathcal{A}_0 \mathcal{B}_0 \rangle} & \color[rgb]{0.501961,0.000000,0.501961}{\langle \mathcal{A}_0 \mathcal{B}_1 \rangle} & \color[rgb]{1.000000,0.501961,0.000000}{\langle  \mathcal{A}_1 \mathcal{B}_0 \rangle}& \langle \mathcal{A}_1 \mathcal{B}_1 \rangle & \color[rgb]{0.400000,0.400000,1.000000}{v_2} \\

 \textcolor{red}{\langle \mathcal{A}_0 \rangle}  & 1 &  \color[rgb]{0.870364,0.870097,0.000000}{v_1} & \color[rgb]{0.250980,0.501961,0.000000}{\langle \mathcal{A}_0 \mathcal{B}_0 \rangle} &  \color[rgb]{0.501961,0.000000,0.501961}{\langle \mathcal{A}_0 \mathcal{B}_1 \rangle} & \langle \mathcal{A}_1 \rangle & \textcolor{blue}{\langle \mathcal{B}_0 \rangle}
 & \langle \mathcal{B}_1 \rangle & \color[rgb]{0.501961,0.000000,0.000000}{v_3} &\color[rgb]{0.501961,1.000000,0.000000}{v_4}& \color[rgb]{1.000000,0.000000,1.000000}{v_5} \\

\langle \mathcal{A}_1 \rangle & \color[rgb]{0.870364,0.870097,0.000000}{v^{\ast}_1}  & 1  &  \color[rgb]{1.000000,0.501961,0.000000}{\langle  \mathcal{A}_1 \mathcal{B}_0 \rangle} &  \langle \mathcal{A}_1 \mathcal{B}_1 \rangle & \textcolor{red}{v_6} & \color[rgb]{0.501961,0.000000,0.000000}{v^{\ast}_3} & \color[rgb]{0.501961,1.000000,0.000000}{v^{\ast}_4} &\textcolor{blue}{\langle \mathcal{B}_0 \rangle} &  \langle \mathcal{B}_1 \rangle & \color[rgb]{0.501961,0.501961,0.000000}{v_7} \\

\textcolor{blue}{\langle \mathcal{B}_0 \rangle} &  \color[rgb]{0.250980,0.501961,0.000000}{\langle \mathcal{A}_0 \mathcal{B}_0 \rangle}& \color[rgb]{1.000000,0.501961,0.000000}{\langle  \mathcal{A}_1 \mathcal{B}_0 \rangle} & 1 & \color[rgb]{0.400000,0.400000,1.000000}{v_2}  & \color[rgb]{0.501961,0.000000,0.000000}{v_3} &  \textcolor{red}{\langle \mathcal{A}_0 \rangle}  & \color[rgb]{1.000000,0.000000,1.000000}{v_5} & \langle \mathcal{A}_1 \rangle & \color[rgb]{0.501961,0.501961,0.000000}{v_7}  & \langle \mathcal{B}_1 \rangle \\

\langle \mathcal{B}_1 \rangle &   \color[rgb]{0.501961,0.000000,0.501961}{\langle \mathcal{A}_0 \mathcal{B}_1 \rangle} &    \langle \mathcal{A}_1 \mathcal{B}_1 \rangle & \color[rgb]{0.400000,0.400000,1.000000}{v^{\ast}_2}  & 1& \color[rgb]{0.501961,1.000000,0.000000}{v_4} & \color[rgb]{1.000000,0.000000,1.000000}{v^{\ast}_5} & \langle \mathcal{A}_0 \rangle & \color[rgb]{0.501961,0.501961,0.000000}{v^{\ast}_7} & \langle \mathcal{A}_1 \rangle & \textcolor{blue}{v_8} \\

 \color[rgb]{0.870364,0.870097,0.000000}{v^{\ast}_1} & \langle \mathcal{A}_1 \rangle & \textcolor{red}{v^{\ast}_6} & \color[rgb]{0.501961,0.000000,0.000000}{v^{\ast}_3} & \color[rgb]{0.501961,1.000000,0.000000}{v^{\ast}_4} & 1 &   \color[rgb]{1.000000,0.501961,0.000000}{\langle  \mathcal{A}_1 \mathcal{B}_0 \rangle} &   \langle \mathcal{A}_1 \mathcal{B}_1 \rangle &  \color[rgb]{0.250980,0.501961,0.000000}{v_9} & \color[rgb]{0.501961,0.000000,0.501961}{v_{10}}& \color[rgb]{0.000000,0.501961,0.501961}{v_{11}} \\

 \color[rgb]{0.250980,0.501961,0.000000}{\langle \mathcal{A}_0 \mathcal{B}_0 \rangle} & \textcolor{blue}{\langle \mathcal{B}_0 \rangle} &  \color[rgb]{0.501961,0.000000,0.000000}{v_3} &  \textcolor{red}{\langle \mathcal{A}_0 \rangle}  & \color[rgb]{1.000000,0.000000,1.000000}{v_5} & \color[rgb]{1.000000,0.501961,0.000000}{\langle  \mathcal{A}_1 \mathcal{B}_0 \rangle} & 1 & \color[rgb]{0.400000,0.400000,1.000000}{v_2}  & \color[rgb]{0.870364,0.870097,0.000000}{v_1} & \color[rgb]{0.000000,0.501961,0.501961}{v_{12}} & \color[rgb]{0.501961,0.000000,0.501961}{\langle \mathcal{A}_0 \mathcal{B}_1 \rangle} \\

\color[rgb]{0.501961,0.000000,0.501961}{\langle \mathcal{A}_0 \mathcal{B}_1 \rangle} & \langle\mathcal{B}_1 \rangle & \color[rgb]{0.501961,1.000000,0.000000}{v_4} &  \color[rgb]{1.000000,0.000000,1.000000}{v^{\ast}_5} &  \textcolor{red}{\langle \mathcal{A}_0 \rangle} & \langle \mathcal{A}_1 \mathcal{B}_1 \rangle & \color[rgb]{0.400000,0.400000,1.000000}{v^{\ast}_2}  & 1 & \color[rgb]{0.000000,0.501961,0.501961}{v_{13}} &  \color[rgb]{0.870364,0.870097,0.000000}{v_1} &  \color[rgb]{0.250980,0.501961,0.000000}{v_{14}} \\

\color[rgb]{1.000000,0.501961,0.000000}{\langle  \mathcal{A}_1 \mathcal{B}_0 \rangle} & \color[rgb]{0.501961,0.000000,0.000000}{v^{\ast}_3} & \textcolor{blue}{\langle \mathcal{B}_0 \rangle} & \langle \mathcal{A}_1 \rangle & \color[rgb]{0.501961,0.501961,0.000000}{v_7}  &  \color[rgb]{0.250980,0.501961,0.000000}{v^{\ast}_9} &  \color[rgb]{0.870364,0.870097,0.000000}{v^{\ast}_1}  & \color[rgb]{0.000000,0.501961,0.501961}{v^{\ast}_{13}}& 1 & \color[rgb]{0.400000,0.400000,1.000000}{v_2}  & \langle \mathcal{A}_1 \mathcal{B}_1 \rangle \\

\langle \mathcal{A}_1 \mathcal{B}_1 \rangle & \color[rgb]{0.501961,1.000000,0.000000}{v^{\ast}_4} & \langle \mathcal{B}_1 \rangle & \color[rgb]{0.501961,0.501961,0.000000}{v^{\ast}_7} & \langle \mathcal{A}_1 \rangle & \color[rgb]{0.501961,0.000000,0.501961}{v^{\ast}_{10}} & \color[rgb]{0.000000,0.501961,0.501961}{v^{\ast}_{12}} &  \color[rgb]{0.870364,0.870097,0.000000}{v^{\ast}_1}  & \color[rgb]{0.400000,0.400000,1.000000}{v^{\ast}_2}  & 1 & \color[rgb]{1.000000,0.501961,0.000000}{v_{15}} \\

\color[rgb]{0.400000,0.400000,1.000000}{v^{\ast}_2} &  \color[rgb]{1.000000,0.000000,1.000000}{v^{\ast}_5} & \color[rgb]{0.501961,0.501961,0.000000}{v^{\ast}_7} & \langle \mathcal{B}_1 \rangle & \textcolor{blue}{v^{\ast}_8} &  \color[rgb]{0.000000,0.501961,0.501961}{v^{\ast}_{11}}& \color[rgb]{0.501961,0.000000,0.501961}{\langle \mathcal{A}_0 \mathcal{B}_1 \rangle} &  \color[rgb]{0.250980,0.501961,0.000000}{v^{\ast}_{14}} & \langle \mathcal{A}_1 \mathcal{B}_1 \rangle &\color[rgb]{1.000000,0.501961,0.000000}{v^{\ast}_{15}} & 1 \\

\end{array}
\right)
\label{example}
\end{equation}
where we define the following unassigned variables
\begin{equation}
\begin{array}{cccc}

v_1 = \langle \mathcal{A}_0 \mathcal{A}_1 \rangle \, ,& v_2 = \langle \mathcal{B}_0 \mathcal{B}_1 \rangle \, , & v_3 = \langle \mathcal{A}_0 \mathcal{A}_1  \mathcal{B}_0 \rangle \, , & v_4 = \langle \mathcal{A}_0 \mathcal{A}_1  \mathcal{B}_1 \rangle  \, ,\\

v_5 = \langle \mathcal{A}_0 \mathcal{B}_0  \mathcal{B}_1 \rangle \, , & v_6 = \langle \mathcal{A}_1 \mathcal{A}_0  \mathcal{A}_1 \rangle \, , & v_7 = \langle \mathcal{A}_1 \mathcal{B}_0  \mathcal{B}_1 \rangle \, , & v_8 = \langle \mathcal{B}_1 \mathcal{B}_0  \mathcal{B}_1 \rangle \, , \\

v_9 = \langle \mathcal{A}_1 \mathcal{A}_0  \mathcal{A}_1  \mathcal{B}_0 \rangle \, , & v_{10} = \langle \mathcal{A}_1 \mathcal{A}_0  \mathcal{A}_1  \mathcal{B}_1 \rangle \, , & v_{11} = \langle \mathcal{A}_1 \mathcal{A}_0  \mathcal{B}_0  \mathcal{B}_1 \rangle \, , & v_{12} = \langle \mathcal{A}_0 \mathcal{A}_1  \mathcal{B}_0  \mathcal{B}_1 \rangle \, , \\

v_{13} = \langle \mathcal{A}_0 \mathcal{A}_1  \mathcal{B}_1  \mathcal{B}_0 \rangle \, , & v_{14} = \langle \mathcal{A}_0 \mathcal{B}_1  \mathcal{B}_0  \mathcal{B}_1 \rangle \, , & v_{15} = \langle \mathcal{A}_1 \mathcal{B}_1  \mathcal{B}_0  \mathcal{B}_1 \rangle \, .\\

\end{array}
\end{equation}

\end{widetext}

Now, if we further impose commutativity of all the measurements,
namely $\left[ \mathcal{A}_0 ,\mathcal{A}_1 \right]  = 0$, $\left[
\mathcal{B}_0 ,\mathcal{B}_1 \right]  = 0$, the corresponding
linear constraints reduce the number of variables. Explicitly, one
gets $v^{\ast}_i = v_i$ for any $i = 1,{\ldots},15$, and also

\begin{equation}
\begin{array}{ccc}
v_6 = \langle \mathcal{A}_0 \rangle \, , & v_8 = \langle \mathcal{B}_0 \rangle \, ,&  v_9 = v_{14} = \langle \mathcal{A}_0 \mathcal{B}_0 \rangle \, ,  \\
v_{10} = \langle \mathcal{A}_0 \mathcal{B}_1 \rangle \, , & v_{15} = \langle \mathcal{A}_1 \mathcal{B}_0 \rangle \, , & v_{11} = v_{12} = v_{13} \, .\\
\end{array}
\label{additional}
\end{equation}

For a visual representation, the variables that become identical because of the commutativity constraints are represented by the same color in (\ref{example}).
For any set of observed correlations $\lbrace \langle \mathcal{A}_x  \rangle, \langle  \mathcal{B}_y  \rangle, \langle \mathcal{A}_x  \mathcal{B}_y \rangle \rbrace $, testing whether it is local can be done in the following steps: assigning the values to the entries of $\Gamma$ that can be derived from the observed correlations and leaving the remaining terms as variables, then checking whether there is an assignment for such variables such that the matrix is positive semidefinite.

For instance, it is possible to check that any set of correlations
that violates the well-known Clauser-Horne-Shimony-Holt (CHSH)
inequality~\cite{clauser1969proposed}
\begin{equation}
\mathcal{I}_{CHSH} =  \langle \mathcal{A}_0 \mathcal{B}_0 \rangle + \langle \mathcal{A}_0 \mathcal{B}_1 \rangle + \langle \mathcal{A}_1 \mathcal{B}_0 \rangle - \langle \mathcal{A}_1 \mathcal{B}_1 \rangle \leq 2
\label{CHSH}
\end{equation}
is incompatible with a positive semidefinite matrix
(\ref{example}). We stress that a necessary condition to produce correlations
that violate (\ref{CHSH}) is that the measurements performed by
each party does not commute with each other. This shows how
the commutativity constraints imposed in the SDP test are crucial
for the detection of the nonlocality of the observed correlations.

To conclude, we notice that, in this particular
scenario, any set of nonlocal correlations has to violate the CHSH inequality (or
symmetrical equivalent of it)~\cite{brunner2014bell}. Therefore,
it turns out that in this case the second level of the hierarchy
is already capable of detecting any nonlocal correlation. That is, even if in this scenario the hierarchy is expected to converge at level $\nu' = 4$, the second level happens already to be tight to the local set.

%%%%%%%%%%%%%%%%%%%%%%%%%%%%%%%%%%%%%%%%%%%%%%%%%%%%%%%%%%%%%
\section{Geometrical characterization of correlations}
\label{sec:geometrical}

\begin{figure}[t]
\includegraphics[scale=0.3]{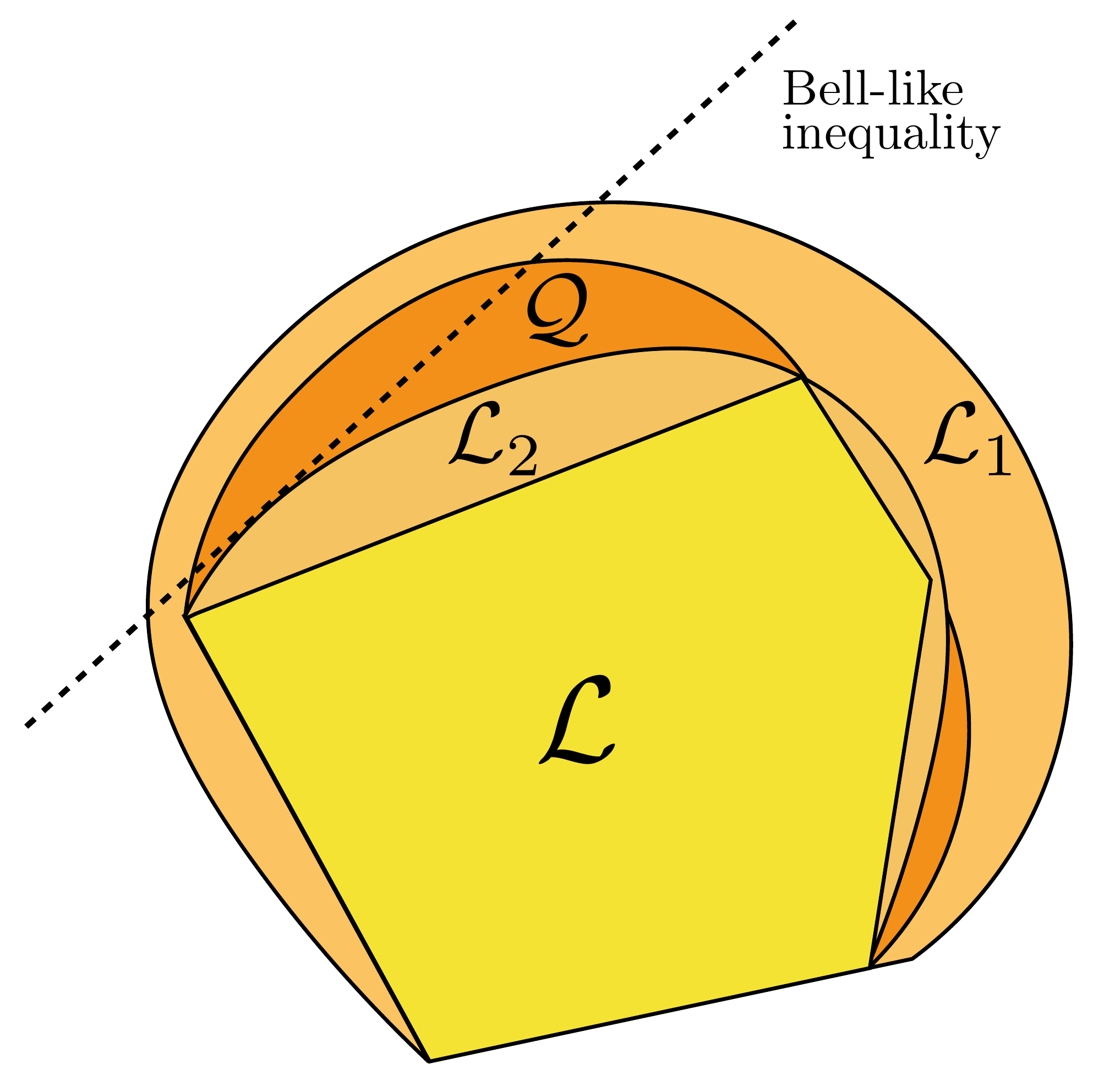}
\caption{Pictorial representation of the sets of correlations, together with our approach to detection of multipartite nonlocality. The $\mathcal{L}$ and $\mathcal{Q}$ sets delimit the local and quantum correlations respectively. As shown here, the first forms a polytope, namely a convex set delimited by a finite amount of extremal points, while the second, despite still being convex, is not a polytope. The light orange sets are the first representatives of the hierarchy $\mathcal{L}_1 \supseteq \mathcal{L}_2 \supseteq {\ldots}\supseteq \mathcal{L}$ approximating the local set from outside. It can be seen that some of the quantum correlations lie outside the $\mathcal{L}_2$, meaning that they are detected as nonlocal from the SDP relaxation at the second level. The dotted line shows a Bell-like inequality that can be obtained by the corresponding dual problem.}
\label{gemetricalpicture}
\end{figure}

Before presenting the applications of our method, we review a geometrical perspective, schematically represented in Figure~\ref{gemetricalpicture}~\cite{brunner2014bell}, that is useful when studying correlations among many different parties.
It is known that the set of local correlations (\ref{local}) defines a polytope, i.e. a convex set with a finite number of extremal points. Such points coincide with the deterministic strategies $D(a_i | x_i, \lambda)$ introduced in (\ref{deter}) and can be easily defined for any multipartite scenario.
As represented in Figure~\ref{gemetricalpicture}, the set of quantum correlations (\ref{quantum}) is strictly bigger than the local set. All the points lying outside the set $\mathcal{L}$ represent nonlocal correlations.

Determining whether some observed correlations are nonlocal corresponds to checking whether they are associated to a point outside the local set. A very simple way to detect nonlocality is by means of Bell inequalities. They are inequalities that are satisfied by any local distribution and geometrically they constitute hyperplanes separating the $\mathcal{L}$ set from the rest of the correlations. Violating a Bell inequality directly implies that the corresponding distribution is nonlocal. However, there can be nonlocal correlations that are not detected by a given inequality, meaning that they fall on the same side of the hyperplane as local correlations.

On the other hand, a very general technique to check if a point belongs to the local set consists in determining if it can be decomposed as a convex combination of its vertices~\cite{zukowski1999strengthening}. Such a question is a typical instance of a linear programing problem, for which there exist algorithms that run in a time that is polynomial in the number of variables~\cite{boyd2004convex}.
Nevertheless, finding a convex decomposition in the multipartite scenario is generally an intractable problem because the number of deterministic strategies grows as $d^{mN}$.
Already in the simplest cases in which each party measures only $m = 2,3$ dicothomic measurements, the best approach currently known stops at $N = 11$ and $N = 7$ respectively~\cite{gondzio2014solving}.

Coming back to the SDP method presented in the previous section, we can now show how the technique can help in overcoming the limitations imposed on the linear program.
Let us define the family of sets $\mathcal{L}_\nu$ as the ones composed by the correlations that are compatible with the moment matrix $\Gamma$ defined by the observables $\mathcal{O}_\nu$ and the additional constraints of commuting measurements.
Given that any local distribution has a quantum representation with commuting measurements, the series $\mathcal{L}_1 \supseteq \mathcal{L}_2 \supseteq {\ldots}\supseteq \mathcal{L}$ defines a hierarchy of sets approximating better and better the local set from outside. In Figure~\ref{gemetricalpicture} we show a schematic representation of the first levels of approximations.

Interestingly, it can be seen that the first level of the hierarchy is not capable of detecting any nonlocal correlations. A simple way to understand it is that, in
the moment matrix generated by $\mathcal{O}_1$, imposing commutativity of the local measurement does not result in any additional constraint in the entries.
A clear example is given by the $N = 2$ case presented in the previous section. The moment matrix corresponding to the first level can be identified with the $5 \times 5$ top-left corner of (\ref{example}). There, the only modification imposed by local commutativity is the condition for the matrix to be real, which can always be assumed when working with quantum correlations.
Therefore, we can say that $\mathcal{L}_1 = \mathcal{Q}_1$, meaning that the first level of our relaxation coincides with the first level of the original NPA, thus resulting in an approximation of the quantum set from outside.

Since we are interested in focusing on the first nontrivial level that allows for nonlocality detection, we then consider $\mathcal{L}_2$. 
We notice that, at this level of the hierarchy, specifying the entries $\Gamma_{ij} = \tr(\mathcal{O}^{\dagger}_i \mathcal{O}_j \rho)$ requires knowledge of up-to-four-body correlators. Moreover, the amount of terms in the set $\mathcal{O}_2$ scales as the number of possible pairs of measurements $M_{x_i}^{a_i}$, that is, as $N^2 m^2 d^2$. This implies that the size of the moment matrix scales only quadratically with the number of parties and measurements, which is much more efficient compared to the exponential dependence $d^{mN}$ of the linear program.
Moreover, since the elements in the moment matrix involve at most four operators, this implies that the number of measurements to be estimated experimentally scales as $N^4 m^4 d^4$.

As mentioned before, checking whether a set of observed correlations belongs to $\mathcal{L}_2$ constitutes a SDP feasibility problem. Since we are addressing approximations of the local set, there will be nonlocal correlations that will fall inside $\mathcal{L}_2$ and that will not be distinguishable from the local correlations. Therefore, our technique can provide only necessary conditions for nonlocality. Nonetheless, we are able to find several examples in which this method is able to successfully detect nonlocal correlations arising from various relevant states, proving that it is not only scalable, but also a powerful method despite being a relaxation.

%%%%%%%%%%%%%%%%%%%%%%%%%%%%%%%%%%%%%%%%%%%%%%%%%%%%%%%%%%%%%%%%%%%%

%%%%%%%%%%%%%%%%%%%%%%%%%%%%%%%%%%%%%%%%%%%%%%%%%%%%%%%%%%%%%%%%%%%

\section{Applications}
\label{sec:results}

The goal of this section is to show that the SDP relaxation can be successfully employed for detection of nonlocality arising from a broad range of quantum states. We focus particularly in exploring the efficient scaling of the method in terms of number of particles. 
To generate the SDP relaxations, we use the software Ncpol2sdpa~\cite{algorithm2015wittek}, and we solve the SDPs with Mosek~\footnote{Available at \url{http://www.mosek.com/}}.

We collect evidence that, from a computational point of view, the main limiting factor of the technique is not time but the amount of memory required to store the moment matrix. Indeed, the longest time that is taken to run one of the codes amounts to approximately $9$ h \footnote{The code used is available under an open source license at \url{https://github.com/FlavioBaccari/Hierarchy-for-nonlocality-detection}}.
Despite the memory limitation, the SDP technique allows us to consider multipartite scenarios that cannot be dealt with in the standard linear program approach to check locality. Indeed, for the scenarios with $m = 2,3$, we are able to detect nonlocality for systems of up to $N = 29$ and $N = 15$ respectively, thus overcoming the current limits of~\cite{gondzio2014solving}.

In the following sections, we list the examples of states we consider. Given that we study cases with dichotomic measurements only, we present them in the expectation value form $\lbrace M^{(i)}_{x_i} \rbrace$.

\subsection*{$W$ state}

As a first case, we analyze the Dicke state with a single excitation, also known as the $W$ state, namely

\begin{equation}
| W_N \rangle = \frac{1}{\sqrt{N}} (\ket{0 {\ldots}01}+\ket{0{\ldots}10}{\ldots}+\ket{10{\ldots}0}).
\label{w}
\end{equation}

Let us consider the simplest scenario of $m = 2$ dichotomic measurements per party, where each observer performs the same two measurements; that is, $M_0^{(i)} = \sigma_x$ and $M_1^{(i)} = \sigma_z$ for all $i = 1,{\ldots},N$.
We are able to show that the obtained probability distribution is detected as nonlocal at level $\mathcal{L}_2$ for $N \leq 29$. 
We recall that in this scenario the complexity of this test scales as $O(N^4)$, in terms of both elements to assign in the moment matrix and measurements to implement experimentally.

We also study the robustness of our technique to white noise, 

\begin{equation}
\rho_N (p) = (1 - p) |W_N \rangle \langle W_N | + p \frac{\mathbb{1}_N}{2^N}
\end{equation}
where $0\leq p\leq 1$ and $\mathbb{1}_N$ represents the identity operator acting on the space of $N$ qubits.
We estimate numerically the maximal value of $p$, referred to as $p_{max}$, for which the given correlations are still nonlocal according to the SDP criterion.
Table \ref{wnoise} reports the resulting values as a function of the number of parties. While the robustness to noise decreases with the number of parties, the method tolerates realistic amounts of noise, always larger than $6\%$, for all the tested configurations.

\begin{table}[h!]
\centering
\begin{tabular}{|c|c||c|c||c|c|}
\hline
$N$ & $p_{max}$ & $N$ & $p_{max}$ & $N$ & $p_{max}$ \\
\hline
5     &    0.295    & 14  & 0.141    & 23 & 0.083  \\
6     &     0.296  &  15  & 0.131    & 24 & 0.079 \\
7     &    0.277   &  16  & 0.122    &  25 & 0.076 \\
8     &    0.251   &  17  & 0.114    &  26 & 0.073     \\
9     &    0.225   &  18  & 0.107  &  27 & 0.070 \\
10   &   0.202   &   19  & 0.101  &  28 &  0.068\\
11   &    0.183  &   20  & 0.096    &  29 & 0.065   \\
12   &    0.167  &   21  & 0.091     && \\
13   &    0.153  &   22  & 0.087    &&\\
\hline
\end{tabular}
\caption{Robustness of nonlocality to white noise in the case of the $W$ state, reported as a function of $N$.}
\label{wnoise}
\end{table}

Finally, in order to study the robustness of the proposed test with respect to the choice of measurements, we also consider a situation where the parties are not able to fully align their measurements and choose randomly two orthogonal measurements \cite{wallman2011generating}. More precisely, we assume that $M_0^{(i)} = \vec{x}_0^{(i)} \cdot \vec{\sigma}$ and  $M_1^{(i)} = \vec{x}_1^{(i)} \cdot \vec{\sigma}$, where $\vec{\sigma} = ( \sigma_x, \sigma_y, \sigma_z ) $ and $ \vec{x}_0^{(i)},  \vec{x}_1^{(i)}$ are vectors chosen uniformly at random, with the only constraint of being orthogonal; namely $\vec{x}_0^{(i)} \cdot  \vec{x}_1^{(i)} = 0$ for all $i = 1,{\ldots},N$.
We calculate numerically the probability $p_{NL}$ for the corresponding correlations to be detected as nonlocal at the second level of the relaxation. To estimate $p_{NL}$, we compute the fraction $N_{NL}/N_r$ of $N_{NL}$ nonlocal distributions obtained over a total of $N_r = 1000$ rounds.
The corresponding results are reported in the following table as a function of $N$.

\begin{table}[h!]
\centering
\begin{tabular}{|c|c||c|c|}
\hline
$N$ & $p_{NL}$ & $N$ & $p_{NL}$ \\
\hline
3  &  50.2 \% & 7 & 21.0 \% \\
4 &  44.4 \%  & 8 & 12.8 \% \\
5 &  38.4 \%  & 9 &  6.3 \%\\
6 &  28.8 \% & 10 & 2.7 \% \\
\hline
\end{tabular}

\end{table}

The results for random measurements also exemplify one of the advantages of our approach with respect to previous entanglement detection schemes. Given some observed correlations, our test can be run and sometimes detects whether the correlations are nonlocal and therefore come from an entangled state. To our understanding, reaching similar conclusions using entanglement witnesses or other entanglement criteria is much harder, as they require solving optimization problems involving $N$-qubit mixed states.

\subsection*{$GHZ$ state}\label{GHZ}

Another well-studied multipartite state is the Greenberger-Horne-Zeilinger (GHZ) state, given by
\begin{equation}
| GHZ_N \rangle = \frac{1}{\sqrt{2}} \left ( | 0 \rangle^{\otimes N} +  | 1 \rangle^{\otimes N} \right).
\label{GHZ}
\end{equation} Contrarily to the $W$ state, such a state is not suited for detection of nonlocality with few-body correlations because all the $k$-body distributions arising from measurements on (\ref{GHZ}) are the same as those obtained by measuring the separable mixed state $\frac{1}{2} ( |0 \rangle \langle 0 |^{\otimes k} + |1 \rangle \langle 1 |^{\otimes k} )$.
Therefore, in order to apply our nonlocality detection method to the $GHZ$ state we need to involve at least one full-body term. 

The solutions we present are inspired by the self-testing scheme for graph states introduced in~\cite{mckague2011self}: the first scenario involves $m = 3$ dichotomic measurements per party; namely $M_0^{(i)} = \sigma_x$, $M_1^{(i)} = \sigma_d = \frac{1}{\sqrt{2}} (\sigma_x + \sigma_z)$ and $M_2^{(i)} = \sigma_z$ for all $i = 1,{\ldots},N$. To introduce full-body correlators in the SDP we define the set $\mathcal{O}_{mix} = \lbrace \mathcal{O}_2 , \langle M_0^{(1)} M_0^{(2)} {\ldots} M_0^{(N)} \rangle, \langle M_1^{(1)} M_0^{(2)} {\ldots} M_0^{(N)} \rangle \rbrace $. The moment matrix corresponding to such set represents a mixed level of the relaxation, containing also two full-body correlators in the entries. However, since the number of added columns and rows is fixed to $2$ for any $N$, this level is basically equivalent to level $\mathcal{L}_2$. Therefore, we preserve the efficient $O(N^4)$ scaling  with the number of parties of elements in the moment matrix and measurements to implement.

By numerically solving the SDP associated to this mixed level of the hierarchy we are able to confirm nonlocality of the correlations arising from the $GHZ$ state and the given measurement for up to $N \leq 15$ parties. Moreover, we check that the number of full-body values that is necessary to assign is constant for any of the considered $N$, coinciding with the two correlators $\langle M_0^{(1)} M_0^{(2)} {\ldots} M_0^{(N)} \rangle$ and $\langle M_1^{(1)} M_0^{(2)} {\ldots} M_0^{(N)} \rangle \ $. Lastly, we estimate that the robustness to noise in this case does not depend on $N$ and it amounts to $p_{max} \approx 0.17$.

As a second scenario, we also notice that one can produce nonlocal correlations from the $GHZ$ at the level $\mathcal{O}_{mix}$ by considering $m = 2$ measurement choices only. Indeed, if ones considers $M_0^{(i)} = \sigma_x$, $M_1^{(i)} = \sigma_d $, the resulting correlations are detected as nonlocal for any $N \leq 28$ (the fact that we are not able to reach $N = 29$ is due to the mixed level of the relaxations, which results in a bigger matrix compared the scenario for the $W$ state).
Table (\ref{ghznoise}) shows the corresponding robustness to noise, computed in the same way as for the $W$ state. For both configurations, the noise robustness of our scheme in detecting GHZ states seems to saturate for large $N$ even if the computational (and experimental) effort scales polynomially.

\begin{table}[h!!]
\centering
\begin{tabular}{|c|c||c|c||c|c|}
\hline
$N$ & $p_{max}$ & $N$ & $p_{max}$ & $N$ & $p_{max}$ \\
\hline
5     &    0.107    & 14  & 0.135    & 23 & 0.145  \\
6     &     0.112  &  15  & 0.137    & 24 & 0.146 \\
7     &    0.116  &  16  & 0.138    &  25 & 0.147 \\
8     &    0.120  &  17  & 0.140    &  26 & 0.147  \\
9     &    0.123  &  18  & 0.141  &  27 & 0.148 \\
10   &    0.127   &   19  & 0.142  &  28 &  0.148 \\
11   &    0.129  &   20  & 0.143    &  &  \\
12   &    0.132  &   21  & 0.144     && \\
13   &    0.134  &   22  & 0.145    &&\\
\hline
\end{tabular}
\caption{Robustness of nonlocality to white noise in the case of the $GHZ$ state and $2$ dicothomic measurements per party, reported as a function of $N$.}
\label{ghznoise}
\end{table}

\subsection*{Graph states}

Graph states~\cite{hein2006entanglement} constitute another important family of multipartite entangled states. Such states are defined as follows: consider a graph $G$, i.e. a set of $N$ vertices labeled by $i$ connected by some edges $\mathcal{E}_{ij}$ connecting the vertices $i$ and $j$. We associate a qubit system in the state $\ket{+}_i$ for each edge $i$ and apply a control-$Z$ grate $CZ_{ij}=\text{diag}(1,1,1,-1)$ to every pair of qubits $i$ and $j$ that are linked according to the graph $G$.

\begin{figure}[t!]
\includegraphics[scale=0.4]{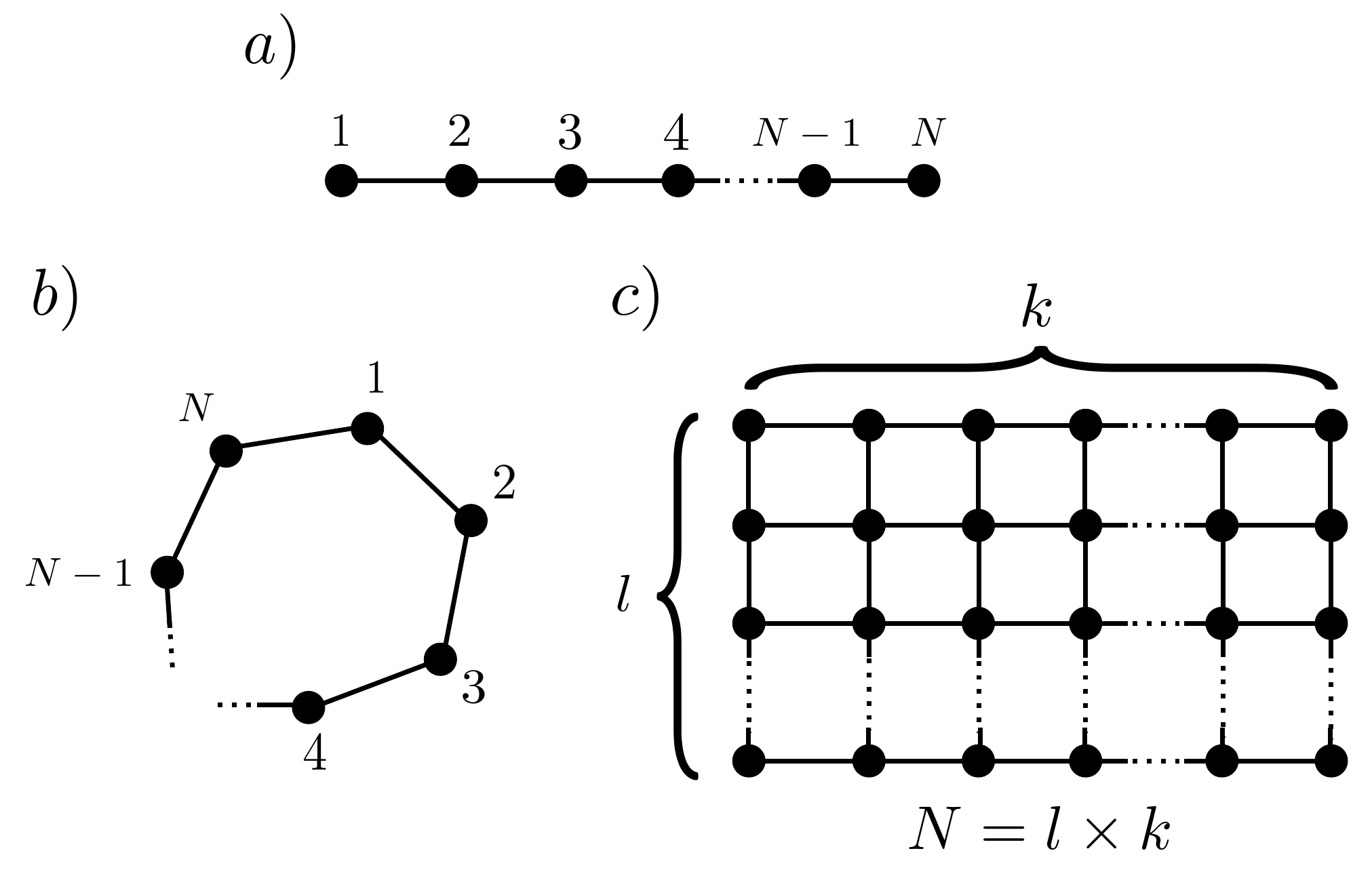}
\caption{Representatives of the graphs associated to the classes of states that have been studied with the SDP method: a) Linear graph states. b) Loop graph states. c) 2D cluster states.}
\label{graphpicture}
\end{figure}

We notice that $GHZ$ state is also a graph state, associated to the so-called star graph. However, due to its particular relevance in quantum information, we prefer to treat its case in the previous section.
Here we consider some other exemplary graph states such as the 1D and 2D cluster states and the loop graph state illustrated in Figure~\ref{graphpicture}. Inspired by the self-testing scheme in~\cite{mckague2011self}, we consider that each party applies three measurements given by $\sigma_x , \sigma_z$ and $\sigma_d$. We are able to detect nonlocality in the obtained correlations at level $\mathcal{L}_2$ for states involving up to $N = 15$ qubits. Again, the method at this level scales as $N^4$.
%%%%%%%%%%%%%%%%%%%%%%%%%%%%%%%%%%%%%%%%%%%%%%%%%%%%%%%%%%%%%%%%%%%%

%%%%%%%%%%%%%%%%%%%%%%%%%%%%%%%%%%%%%%%%%%%%%%%%%%%%%%%%%%%%%%%%%%%

Interestingly, our approach for the detection of nonlocal correlations generated by graph states shows to be qualitatively different from McKague's scheme in~\cite{mckague2011self}. While the latter requires correlators of an order that depends on the connectivity of the graph (namely, equal to $1$ plus the maximal number of neighbors that each vertex has), our method seems - at least in some cases - to be independent of it.
Indeed, we are able to detect nonlocality with four-body correlators in 2D cluster states, whose connectivity would imply five-body correlators for the self-testing scheme.

\section{Explicit Bell inequalities}\label{sec:bell}

Another nice property of our nonlocality criterion comes from the fact that, as it can be put in a SDP form, it immediately provides a method to find experimentally friendly Bell inequalities involving a subset of all possible measurements. In fact, it turns out that the the SDP proposed in Sec.~\ref{sec:method} has a dual formulation that can be interpreted as the optimization of a linear function of the correlations that can be seen as a Bell-like functional, \ie a functional that has a nontrivial bound for all correlations in $\mathcal{L}_k$~\citep{navascues2008convergent} (see Appendix~\ref{app:details} for details). Thus, if a set of correlations is found to be nonlocal, then the solution of the SDP provides a Bell inequality that is satisfied by correlations in $\mathcal{L}_{\nu}$ and that is violated by the tested correlations. Importantly, this Bell inequality can further be used to test other sets of correlations.

By using the two sets of correlations obtained by measuring 3 dicothomic observables per party in the GHZ state we are able to find the following Bell inequality:

\begin{equation}
\begin{split}
\mathcal{I}_{mix}^{3} =  \sum_{i = 2}^{N} \langle M_{1}^{(1)}  M_{2}^{(i)} \rangle - \sum_{i = 2}^{N} \langle M_{0}^{(1)}  M_{2}^{(i)} \rangle + \\ + (N-1) \langle M_{0}^{(1)}  M_{0}^{(2)} {\ldots}  M_{0}^{(N)} \rangle + \\ + (N-1) \langle M_{1}^{(1)}  M_{0}^{(2)} {\ldots}  M_{0}^{(N)} \rangle  \leq 2(N-1)
\end{split}
\label{ineq3}
\end{equation}
Numerically, we could certify the validity of this inequality for up to $N\leq15$. Moreover, in principle the bound of $\beta_C = 2(N-1)$ is only guaranteed to be satisfied by correlations in $\mathcal{L}_{mix}$. However, motivated by the obtained numerical insight, we could prove that this bound actually coincides with the true local bound, and therefore, \eqref{ineq3} is a valid Bell inequality for all $N$  (for all the analytical proofs regarding this section, see Appendix~\ref{app:ineq}). This shows that, at least in this instance, the $\mathcal{L}_{mix}$ defined by the SDP relaxation associated to $\mathcal{O}_{mix}$ is tight to the local set.

It is also easy to show that (\ref{ineq3}) is violated by the $GHZ$ state and the previously introduced choice of measurements. In particular, the value reached is $\mathcal{I}^3_{GHZ} = (1 + \sqrt{2}) (N-1)$ for any $N$.
Given that both the local bound and the violation scale linearly with $N$, the robustness of nonlocality to white noise is constant and amounts to $p_{max} = \frac{\sqrt{2} -1 }{\sqrt{2} + 1} \approx 0.174$. We note that this results is in agreement with what is achieved numerically with the SDP for up to $N = 15$.

Similarly, we also find the following Bell inequality by using the set of correlations involving only two measurements per party described for the GHZ state:

\begin{equation}
\begin{split}
\mathcal{I}_{mix}^{2} =  \sum_{i = 2}^{N} \langle M_{1}^{(1)}  M_{1}^{(i)} \rangle - \sum_{i = 2}^{N} \langle M_{0}^{(1)}  M_{1}^{(i)} \rangle + \\ + (N-1) \langle M_{0}^{(1)}  M_{0}^{(2)} {\ldots}  M_{0}^{(N)} \rangle + \\ + (N-1) \langle M_{1}^{(1)}  M_{0}^{(2)} {\ldots}  M_{0}^{(N)} \rangle  \leq 2(N-1)
\end{split}
\label{ineq2}
\end{equation}
Once more, although this inequality is found numerically for up to $N\leq28$ we prove that it is valid for any $N$. Moreover the bound $\beta_C=2(N-1)$ is not only valid for correlation in $\mathcal{L}_{mix}$ but for any local set of correlations. The $GHZ$ state and the given measurements result in a violation of $\mathcal{I}_{GHZ}^{2} = \frac{3 + \sqrt{2}}{2}$. Given that in this case the relative violation is lower, we also have a lower robustness to noise, coinciding with $p_{max} = \frac{\sqrt{2} - 1}{\sqrt{2} +3} \approx 0.09$ for any $N$.
We notice that this value is different from the ones reported in Table (\ref{ghznoise}). The reason is that, to derive inequality (\ref{ineq2}) from the dual, we restrict to assign only the values of the two-body correlations and the two full-body ones. On the other hand, the results in Table (\ref{ghznoise}) also take into account the assignment of the three- and four-body correlators, showing that this additional knowledge helps in improving the robustness to noise.

As a final remark, we stress that the measurement settings considered to derive an inequality from the dual might not be the optimal ones. For instance, we are able to identify different measurement choices for the case of (\ref{ineq2}) that lead to a higher violation of such an inequality, hence resulting also in a better robustness to noise (see Appendix~\ref{app:ineq} for details).

\section{Discussion}

We introduce a technique for efficient device-independent entanglement detection for multipartite quantum systems. It relies on a hierarchy of necessary conditions for nonlocality in the observed correlations. By focusing on the second level of the hierarchy, we consider a test that requires knowledge of up to four-body correlators only. We show that it can be successfully applied to detect entanglement of many physically relevant states, such as the $W$, the $GHZ$  and the graph states.
Besides being suitable for experimental implementation, our technique also has an efficient scaling in terms of computational requirements, given that the number of variables involved grows polynomially with $N$. This allows us to overcome the limitation of the currently known methods and to detect entanglement for states of up to few tens of particles.
Moreover, the proposed technique has a completely general approach and it can be applied to any set of observed correlations. This makes it particularly relevant for the detection of new classes of multipartite entangled states.

We note that our techniques can also be used as a semidefinite constraints to impose locality. Consider, for instance, a linear function $f$ of the observed correlations. One could find an upper bound on the value of this function over local correlations by maximizing it under the constraint that the moment matrix $\Gamma$ is positive semidefinite. A particular example could be to take $f$ to be a Bell polynomial. Thus this approach would find a bound $f\leq \beta_C$ satisfied by all local correlations.

As a future question in this direction, it would be interesting to study how accurate is the approximation of the local set of correlations provided by the second level of the hierarchy. In some of the scenario that we consider the approximation is actually tight, but this is not generally the case. A possible approach could be to compare the local bound of some known Bell inequalities with that resulting from the hierarchy.

Furthermore, we notice that the second level of the hierarchy also has an efficient scaling with the number of measurements performed by the parties. This would allow us to inquire whether an increasing number of measurement choices can provide an advantage for entanglement detection in multipartite systems.

Lastly, we believe that the present techniques can be readily applied in current state-of-the-art experiments. For instance, experiments composed by up to 7 ions have demonstrated nonlocality using an exponentially increasing number of full correlators \cite{lanyon2014experimental}. Moreover, recent experiments have produced GHZ-like states in systems composed by 14 ions \cite{monz201114qubit} and 10 photons \cite{wang2016experimental,chen2017observation} with visibilities within the range required to observe a violation of the Bell inequalities presented here. We notice, however, that the measurements required to certify the presence of nonlocal correlations using our approach are different from the ones reported in these works.

\label{sec:conclusions}

\section{Acknowledgements}
\label{sec:acknowledgements}

The authors thank J. Kaniewski  and the referees for helpful comments. D.C. thanks R. Chaves for discussions and hospitality at the International Institute of Physics (Natal-Brazil).
This work was supported by the ERC CoG grant QITBOX, Spanish MINECO (Severo Ochoa grant SEV-2015-0522, a  Ram\'on y Cajal fellowship, a Severo Ochoa PhD fellowship, QIBEQI FIS2016-80773-P and FOQUS FIS2013-46768-P), the AXA Chair in Quantum Information Science, the Generalitat de Catalunya (SGR875 and CERCA Program) and Fundacion Cellex.
The authors also acknowledge the computational resources granted by the High Performance Computing Center North
(SNIC 2016/1-320).

\bibliography{Biblio}

\begin{thebibliography}{42}%
\makeatletter
\providecommand \@ifxundefined [1]{%
 \@ifx{#1\undefined}
}%
\providecommand \@ifnum [1]{%
 \ifnum #1\expandafter \@firstoftwo
 \else \expandafter \@secondoftwo
 \fi
}%
\providecommand \@ifx [1]{%
 \ifx #1\expandafter \@firstoftwo
 \else \expandafter \@secondoftwo
 \fi
}%
\providecommand \natexlab [1]{#1}%
\providecommand \enquote  [1]{``#1''}%
\providecommand \bibnamefont  [1]{#1}%
\providecommand \bibfnamefont [1]{#1}%
\providecommand \citenamefont [1]{#1}%
\providecommand \href@noop [0]{\@secondoftwo}%
\providecommand \href [0]{\begingroup \@sanitize@url \@href}%
\providecommand \@href[1]{\@@startlink{#1}\@@href}%
\providecommand \@@href[1]{\endgroup#1\@@endlink}%
\providecommand \@sanitize@url [0]{\catcode `\\12\catcode `\$12\catcode
  `\&12\catcode `\#12\catcode `\^12\catcode `\_12\catcode `\%12\relax}%
\providecommand \@@startlink[1]{}%
\providecommand \@@endlink[0]{}%
\providecommand \url  [0]{\begingroup\@sanitize@url \@url }%
\providecommand \@url [1]{\endgroup\@href {#1}{\urlprefix }}%
\providecommand \urlprefix  [0]{URL }%
\providecommand \Eprint [0]{\href }%
\providecommand \doibase [0]{http://dx.doi.org/}%
\providecommand \selectlanguage [0]{\@gobble}%
\providecommand \bibinfo  [0]{\@secondoftwo}%
\providecommand \bibfield  [0]{\@secondoftwo}%
\providecommand \translation [1]{[#1]}%
\providecommand \BibitemOpen [0]{}%
\providecommand \bibitemStop [0]{}%
\providecommand \bibitemNoStop [0]{.\EOS\space}%
\providecommand \EOS [0]{\spacefactor3000\relax}%
\providecommand \BibitemShut  [1]{\csname bibitem#1\endcsname}%
\let\auto@bib@innerbib\@empty
%</preamble>
\bibitem [{\citenamefont {Bennett}\ \emph {et~al.}(1993)\citenamefont
  {Bennett}, \citenamefont {Brassard}, \citenamefont {Cr\'epeau}, \citenamefont
  {Jozsa}, \citenamefont {Peres},\ and\ \citenamefont
  {Wootters}}]{bennett1993teleporting}%
  \BibitemOpen
  \bibfield  {author} {\bibinfo {author} {\bibfnamefont {Charles~H.}\
  \bibnamefont {Bennett}}, \bibinfo {author} {\bibfnamefont {Gilles}\
  \bibnamefont {Brassard}}, \bibinfo {author} {\bibfnamefont {Claude}\
  \bibnamefont {Cr\'epeau}}, \bibinfo {author} {\bibfnamefont {Richard}\
  \bibnamefont {Jozsa}}, \bibinfo {author} {\bibfnamefont {Asher}\ \bibnamefont
  {Peres}}, \ and\ \bibinfo {author} {\bibfnamefont {William~K.}\ \bibnamefont
  {Wootters}},\ }\bibfield  {title} {\enquote {\bibinfo {title} {Teleporting an
  unknown quantum state via dual classical and einstein-podolsky-rosen
  channels},}\ }\href {\doibase 10.1103/PhysRevLett.70.1895} {\bibfield
  {journal} {\bibinfo  {journal} {Physical Review Letters}\ }\textbf {\bibinfo
  {volume} {70}},\ \bibinfo {pages} {1895--1899} (\bibinfo {year}
  {1993})}\BibitemShut {NoStop}%
\bibitem [{\citenamefont {Ekert}(1991)}]{ekert1991quantum}%
  \BibitemOpen
  \bibfield  {author} {\bibinfo {author} {\bibfnamefont {Artur~K.}\
  \bibnamefont {Ekert}},\ }\bibfield  {title} {\enquote {\bibinfo {title}
  {Quantum cryptography based on {B}ell's theorem},}\ }\href {\doibase
  10.1103/PhysRevLett.67.661} {\bibfield  {journal} {\bibinfo  {journal}
  {Physical Review Letters}\ }\textbf {\bibinfo {volume} {67}},\ \bibinfo
  {pages} {661--663} (\bibinfo {year} {1991})}\BibitemShut {NoStop}%
\bibitem [{\citenamefont {Briegel}\ \emph {et~al.}(2009)\citenamefont
  {Briegel}, \citenamefont {Browne}, \citenamefont {D{\"{u}}r}, \citenamefont
  {Raussendorf},\ and\ \citenamefont {{Van den
  Nest}}}]{briegel2009measurement}%
  \BibitemOpen
  \bibfield  {author} {\bibinfo {author} {\bibfnamefont {H~J}\ \bibnamefont
  {Briegel}}, \bibinfo {author} {\bibfnamefont {D~E}\ \bibnamefont {Browne}},
  \bibinfo {author} {\bibfnamefont {W}~\bibnamefont {D{\"{u}}r}}, \bibinfo
  {author} {\bibfnamefont {R}~\bibnamefont {Raussendorf}}, \ and\ \bibinfo
  {author} {\bibfnamefont {M}~\bibnamefont {{Van den Nest}}},\ }\bibfield
  {title} {\enquote {\bibinfo {title} {{Measurement-based quantum
  computation}},}\ }\href {\doibase 10.1038/nphys1157} {\bibfield  {journal}
  {\bibinfo  {journal} {Nature Physics}\ }\textbf {\bibinfo {volume} {5}},\
  \bibinfo {pages} {19--26} (\bibinfo {year} {2009})}\BibitemShut {NoStop}%
\bibitem [{\citenamefont {Giovannetti}\ \emph {et~al.}(2004)\citenamefont
  {Giovannetti}, \citenamefont {Lloyd},\ and\ \citenamefont
  {Maccone}}]{giovannetti2004quantum}%
  \BibitemOpen
  \bibfield  {author} {\bibinfo {author} {\bibfnamefont {Vittorio}\
  \bibnamefont {Giovannetti}}, \bibinfo {author} {\bibfnamefont {Seth}\
  \bibnamefont {Lloyd}}, \ and\ \bibinfo {author} {\bibfnamefont {Lorenzo}\
  \bibnamefont {Maccone}},\ }\bibfield  {title} {\enquote {\bibinfo {title}
  {{Quantum-enhanced measurements: beating the standard quantum limit.}}}\
  }\href {\doibase 10.1126/science.1104149} {\bibfield  {journal} {\bibinfo
  {journal} {Science}\ }\textbf {\bibinfo {volume} {306}},\ \bibinfo {pages}
  {1330--6} (\bibinfo {year} {2004})}\BibitemShut {NoStop}%
\bibitem [{\citenamefont {Horodecki}\ \emph {et~al.}(2009)\citenamefont
  {Horodecki}, \citenamefont {Horodecki}, \citenamefont {Horodecki},\ and\
  \citenamefont {Horodecki}}]{horodecki2009entanglement}%
  \BibitemOpen
  \bibfield  {author} {\bibinfo {author} {\bibfnamefont {Ryszard}\ \bibnamefont
  {Horodecki}}, \bibinfo {author} {\bibfnamefont {Pawe\l{}}\ \bibnamefont
  {Horodecki}}, \bibinfo {author} {\bibfnamefont {Micha\l{}}\ \bibnamefont
  {Horodecki}}, \ and\ \bibinfo {author} {\bibfnamefont {Karol}\ \bibnamefont
  {Horodecki}},\ }\bibfield  {title} {\enquote {\bibinfo {title} {Quantum
  entanglement},}\ }\href {\doibase 10.1103/RevModPhys.81.865} {\bibfield
  {journal} {\bibinfo  {journal} {Review of Modern Physics}\ }\textbf {\bibinfo
  {volume} {81}},\ \bibinfo {pages} {865--942} (\bibinfo {year}
  {2009})}\BibitemShut {NoStop}%
\bibitem [{\citenamefont {H{\"a}ffner}\ \emph {et~al.}(2005)\citenamefont
  {H{\"a}ffner}, \citenamefont {H{\"a}nsel}, \citenamefont {Roos},
  \citenamefont {Benhelm}, \citenamefont {Chwalla}, \citenamefont {K{\"o}rber},
  \citenamefont {Rapol}, \citenamefont {Riebe}, \citenamefont {Schmidt},
  \citenamefont {Becher} \emph {et~al.}}]{haffner2005scalable}%
  \BibitemOpen
  \bibfield  {author} {\bibinfo {author} {\bibfnamefont {H}~\bibnamefont
  {H{\"a}ffner}}, \bibinfo {author} {\bibfnamefont {W}~\bibnamefont
  {H{\"a}nsel}}, \bibinfo {author} {\bibfnamefont {CF}~\bibnamefont {Roos}},
  \bibinfo {author} {\bibfnamefont {J}~\bibnamefont {Benhelm}}, \bibinfo
  {author} {\bibfnamefont {M}~\bibnamefont {Chwalla}}, \bibinfo {author}
  {\bibfnamefont {T}~\bibnamefont {K{\"o}rber}}, \bibinfo {author}
  {\bibfnamefont {UD}~\bibnamefont {Rapol}}, \bibinfo {author} {\bibfnamefont
  {M}~\bibnamefont {Riebe}}, \bibinfo {author} {\bibfnamefont {PO}~\bibnamefont
  {Schmidt}}, \bibinfo {author} {\bibfnamefont {C}~\bibnamefont {Becher}},
  \emph {et~al.},\ }\bibfield  {title} {\enquote {\bibinfo {title} {Scalable
  multiparticle entanglement of trapped ions},}\ }\href {\doibase
  10.1038/nature04279} {\bibfield  {journal} {\bibinfo  {journal} {Nature}\
  }\textbf {\bibinfo {volume} {438}},\ \bibinfo {pages} {643--646} (\bibinfo
  {year} {2005})}\BibitemShut {NoStop}%
\bibitem [{\citenamefont {Brand{\~{a}}o}\ and\ \citenamefont
  {Christandl}(2012)}]{brandao2012detection}%
  \BibitemOpen
  \bibfield  {author} {\bibinfo {author} {\bibfnamefont {Fernando G S~L}\
  \bibnamefont {Brand{\~{a}}o}}\ and\ \bibinfo {author} {\bibfnamefont
  {Matthias}\ \bibnamefont {Christandl}},\ }\bibfield  {title} {\enquote
  {\bibinfo {title} {Detection of multiparticle entanglement: Quantifying the
  search for symmetric extensions},}\ }\href {\doibase
  10.1103/PhysRevLett.109.160502} {\bibfield  {journal} {\bibinfo  {journal}
  {Physical Review Letters}\ }\textbf {\bibinfo {volume} {109}},\ \bibinfo
  {pages} {160502} (\bibinfo {year} {2012})}\BibitemShut {NoStop}%
\bibitem [{\citenamefont {Bourennane}\ \emph {et~al.}(2004)\citenamefont
  {Bourennane}, \citenamefont {Eibl}, \citenamefont {Kurtsiefer}, \citenamefont
  {Gaertner}, \citenamefont {Weinfurter}, \citenamefont {G{\"u}hne},
  \citenamefont {Hyllus}, \citenamefont {Bru{\ss}}, \citenamefont
  {Lewenstein},\ and\ \citenamefont {Sanpera}}]{bourennane2004experimental}%
  \BibitemOpen
  \bibfield  {author} {\bibinfo {author} {\bibfnamefont {Mohamed}\ \bibnamefont
  {Bourennane}}, \bibinfo {author} {\bibfnamefont {Manfred}\ \bibnamefont
  {Eibl}}, \bibinfo {author} {\bibfnamefont {Christian}\ \bibnamefont
  {Kurtsiefer}}, \bibinfo {author} {\bibfnamefont {Sascha}\ \bibnamefont
  {Gaertner}}, \bibinfo {author} {\bibfnamefont {Harald}\ \bibnamefont
  {Weinfurter}}, \bibinfo {author} {\bibfnamefont {Otfried}\ \bibnamefont
  {G{\"u}hne}}, \bibinfo {author} {\bibfnamefont {Philipp}\ \bibnamefont
  {Hyllus}}, \bibinfo {author} {\bibfnamefont {Dagmar}\ \bibnamefont
  {Bru{\ss}}}, \bibinfo {author} {\bibfnamefont {Maciej}\ \bibnamefont
  {Lewenstein}}, \ and\ \bibinfo {author} {\bibfnamefont {Anna}\ \bibnamefont
  {Sanpera}},\ }\bibfield  {title} {\enquote {\bibinfo {title} {Experimental
  detection of multipartite entanglement using witness operators},}\ }\href
  {\doibase 10.1103/PhysRevLett.92.087902} {\bibfield  {journal} {\bibinfo
  {journal} {Physical Review Letters}\ }\textbf {\bibinfo {volume} {92}},\
  \bibinfo {pages} {087902} (\bibinfo {year} {2004})}\BibitemShut {NoStop}%
\bibitem [{\citenamefont {T{\'{o}}th}\ and\ \citenamefont
  {G{\"{u}}hne}(2006)}]{toth2006detection}%
  \BibitemOpen
  \bibfield  {author} {\bibinfo {author} {\bibfnamefont {G.}~\bibnamefont
  {T{\'{o}}th}}\ and\ \bibinfo {author} {\bibfnamefont {O.}~\bibnamefont
  {G{\"{u}}hne}},\ }\bibfield  {title} {\enquote {\bibinfo {title} {{Detection
  of multipartite entanglement with two-body correlations}},}\ }\href {\doibase
  10.1007/s00340-005-2057-1} {\bibfield  {journal} {\bibinfo  {journal}
  {Applied Physics B: Lasers and Optics}\ }\textbf {\bibinfo {volume} {82}},\
  \bibinfo {pages} {237--241} (\bibinfo {year} {2006})}\BibitemShut {NoStop}%
\bibitem [{\citenamefont {L\"ucke}\ \emph {et~al.}(2014)\citenamefont
  {L\"ucke}, \citenamefont {Peise}, \citenamefont {Vitagliano}, \citenamefont
  {Arlt}, \citenamefont {Santos}, \citenamefont {T\'oth},\ and\ \citenamefont
  {Klempt}}]{lucke2014detecting}%
  \BibitemOpen
  \bibfield  {author} {\bibinfo {author} {\bibfnamefont {Bernd}\ \bibnamefont
  {L\"ucke}}, \bibinfo {author} {\bibfnamefont {Jan}\ \bibnamefont {Peise}},
  \bibinfo {author} {\bibfnamefont {Giuseppe}\ \bibnamefont {Vitagliano}},
  \bibinfo {author} {\bibfnamefont {Jan}\ \bibnamefont {Arlt}}, \bibinfo
  {author} {\bibfnamefont {Luis}\ \bibnamefont {Santos}}, \bibinfo {author}
  {\bibfnamefont {G\'eza}\ \bibnamefont {T\'oth}}, \ and\ \bibinfo {author}
  {\bibfnamefont {Carsten}\ \bibnamefont {Klempt}},\ }\bibfield  {title}
  {\enquote {\bibinfo {title} {Detecting multiparticle entanglement of {D}icke
  states},}\ }\href {\doibase 10.1103/PhysRevLett.112.155304} {\bibfield
  {journal} {\bibinfo  {journal} {Physical Review Letters}\ }\textbf {\bibinfo
  {volume} {112}},\ \bibinfo {pages} {155304} (\bibinfo {year}
  {2014})}\BibitemShut {NoStop}%
\bibitem [{\citenamefont {Vitagliano}\ \emph {et~al.}(2014)\citenamefont
  {Vitagliano}, \citenamefont {Apellaniz}, \citenamefont {Egusquiza},\ and\
  \citenamefont {T{\'{o}}th}}]{vitagliano2014spin}%
  \BibitemOpen
  \bibfield  {author} {\bibinfo {author} {\bibfnamefont {Giuseppe}\
  \bibnamefont {Vitagliano}}, \bibinfo {author} {\bibfnamefont {Iagoba}\
  \bibnamefont {Apellaniz}}, \bibinfo {author} {\bibfnamefont {I{\~{n}}igo~L.}\
  \bibnamefont {Egusquiza}}, \ and\ \bibinfo {author} {\bibfnamefont
  {G{\'{e}}za}\ \bibnamefont {T{\'{o}}th}},\ }\bibfield  {title} {\enquote
  {\bibinfo {title} {Spin squeezing and entanglement for an arbitrary spin},}\
  }\href {\doibase 10.1103/PhysRevA.89.032307} {\bibfield  {journal} {\bibinfo
  {journal} {Physical Review A}\ }\textbf {\bibinfo {volume} {89}},\ \bibinfo
  {pages} {032307} (\bibinfo {year} {2014})}\BibitemShut {NoStop}%
\bibitem [{\citenamefont {Jungnitsch}\ \emph {et~al.}(2011)\citenamefont
  {Jungnitsch}, \citenamefont {Moroder},\ and\ \citenamefont
  {G{\"{u}}hne}}]{jungnitsch2011taming}%
  \BibitemOpen
  \bibfield  {author} {\bibinfo {author} {\bibfnamefont {Bastian}\ \bibnamefont
  {Jungnitsch}}, \bibinfo {author} {\bibfnamefont {Tobias}\ \bibnamefont
  {Moroder}}, \ and\ \bibinfo {author} {\bibfnamefont {Otfried}\ \bibnamefont
  {G{\"{u}}hne}},\ }\bibfield  {title} {\enquote {\bibinfo {title} {{Taming
  multiparticle entanglement}},}\ }\href {\doibase
  10.1103/PhysRevLett.106.190502} {\bibfield  {journal} {\bibinfo  {journal}
  {Physical Review Letters}\ }\textbf {\bibinfo {volume} {106}},\ \bibinfo
  {pages} {190502} (\bibinfo {year} {2011})}\BibitemShut {NoStop}%
\bibitem [{\citenamefont {Knips}\ \emph {et~al.}(2016)\citenamefont {Knips},
  \citenamefont {Schwemmer}, \citenamefont {Klein}, \citenamefont
  {Wie\ifmmode~\acute{s}\else \'{s}\fi{}niak},\ and\ \citenamefont
  {Weinfurter}}]{knips2014multipartite}%
  \BibitemOpen
  \bibfield  {author} {\bibinfo {author} {\bibfnamefont {Lukas}\ \bibnamefont
  {Knips}}, \bibinfo {author} {\bibfnamefont {Christian}\ \bibnamefont
  {Schwemmer}}, \bibinfo {author} {\bibfnamefont {Nico}\ \bibnamefont {Klein}},
  \bibinfo {author} {\bibfnamefont {Marcin}\ \bibnamefont
  {Wie\ifmmode~\acute{s}\else \'{s}\fi{}niak}}, \ and\ \bibinfo {author}
  {\bibfnamefont {Harald}\ \bibnamefont {Weinfurter}},\ }\bibfield  {title}
  {\enquote {\bibinfo {title} {Multipartite entanglement detection with minimal
  effort},}\ }\href {\doibase 10.1103/PhysRevLett.117.210504} {\bibfield
  {journal} {\bibinfo  {journal} {Physical Review Letters}\ }\textbf {\bibinfo
  {volume} {117}},\ \bibinfo {pages} {210504} (\bibinfo {year}
  {2016})}\BibitemShut {NoStop}%
\bibitem [{\citenamefont {Apellaniz}\ \emph {et~al.}(2017)\citenamefont
  {Apellaniz}, \citenamefont {Kleinmann}, \citenamefont {G\"uhne},\ and\
  \citenamefont {T\'oth}}]{apellaniz2015optimal}%
  \BibitemOpen
  \bibfield  {author} {\bibinfo {author} {\bibfnamefont {Iagoba}\ \bibnamefont
  {Apellaniz}}, \bibinfo {author} {\bibfnamefont {Matthias}\ \bibnamefont
  {Kleinmann}}, \bibinfo {author} {\bibfnamefont {Otfried}\ \bibnamefont
  {G\"uhne}}, \ and\ \bibinfo {author} {\bibfnamefont {G\'eza}\ \bibnamefont
  {T\'oth}},\ }\bibfield  {title} {\enquote {\bibinfo {title} {Optimal
  witnessing of the quantum fisher information with few measurements},}\ }\href
  {\doibase 10.1103/PhysRevA.95.032330} {\bibfield  {journal} {\bibinfo
  {journal} {Physical Review A}\ }\textbf {\bibinfo {volume} {95}},\ \bibinfo
  {pages} {032330} (\bibinfo {year} {2017})}\BibitemShut {NoStop}%
\bibitem [{\citenamefont {Bell}(1964)}]{bell1964on}%
  \BibitemOpen
  \bibfield  {author} {\bibinfo {author} {\bibfnamefont {John~Stewart}\
  \bibnamefont {Bell}},\ }\bibfield  {title} {\enquote {\bibinfo {title} {On
  the {E}instein-{P}odolsky-{R}osen paradox},}\ }\href
  {https://cds.cern.ch/record/111654} {\bibfield  {journal} {\bibinfo
  {journal} {Physics}\ }\textbf {\bibinfo {volume} {1}},\ \bibinfo {pages}
  {195--200} (\bibinfo {year} {1964})}\BibitemShut {NoStop}%
\bibitem [{\citenamefont {Brunner}\ \emph {et~al.}(2014)\citenamefont
  {Brunner}, \citenamefont {Cavalcanti}, \citenamefont {Pironio}, \citenamefont
  {Scarani},\ and\ \citenamefont {Wehner}}]{brunner2014bell}%
  \BibitemOpen
  \bibfield  {author} {\bibinfo {author} {\bibfnamefont {Nicolas}\ \bibnamefont
  {Brunner}}, \bibinfo {author} {\bibfnamefont {Daniel}\ \bibnamefont
  {Cavalcanti}}, \bibinfo {author} {\bibfnamefont {Stefano}\ \bibnamefont
  {Pironio}}, \bibinfo {author} {\bibfnamefont {Valerio}\ \bibnamefont
  {Scarani}}, \ and\ \bibinfo {author} {\bibfnamefont {Stephanie}\ \bibnamefont
  {Wehner}},\ }\bibfield  {title} {\enquote {\bibinfo {title} {Bell
  nonlocality},}\ }\href {\doibase 10.1103/RevModPhys.86.419} {\bibfield
  {journal} {\bibinfo  {journal} {Reviews of Modern Physics}\ }\textbf
  {\bibinfo {volume} {86}},\ \bibinfo {pages} {419} (\bibinfo {year}
  {2014})}\BibitemShut {NoStop}%
\bibitem [{\citenamefont {Zukowski}\ \emph {et~al.}(1999)\citenamefont
  {Zukowski}, \citenamefont {Kaszlikowski}, \citenamefont {Baturo},\ and\
  \citenamefont {Larsson}}]{zukowski1999strengthening}%
  \BibitemOpen
  \bibfield  {author} {\bibinfo {author} {\bibfnamefont {Marek}\ \bibnamefont
  {Zukowski}}, \bibinfo {author} {\bibfnamefont {Dagomir}\ \bibnamefont
  {Kaszlikowski}}, \bibinfo {author} {\bibfnamefont {Adam}\ \bibnamefont
  {Baturo}}, \ and\ \bibinfo {author} {\bibfnamefont {Jan-Ake}\ \bibnamefont
  {Larsson}},\ }\bibfield  {title} {\enquote {\bibinfo {title} {Strengthening
  the {B}ell theorem: conditions to falsify local realism in an experiment},}\
  }\href {https://arxiv.org/abs/quant-ph/9910058} {\bibfield  {journal}
  {\bibinfo  {journal} {arXiv quant-ph/9910058}\ } (\bibinfo {year}
  {1999})}\BibitemShut {NoStop}%
\bibitem [{\citenamefont {Fine}(1982)}]{fine1982hidden}%
  \BibitemOpen
  \bibfield  {author} {\bibinfo {author} {\bibfnamefont {Arthur}\ \bibnamefont
  {Fine}},\ }\bibfield  {title} {\enquote {\bibinfo {title} {Hidden variables,
  joint probability, and the {B}ell inequalities},}\ }\href {\doibase
  10.1103/PhysRevLett.48.291} {\bibfield  {journal} {\bibinfo  {journal}
  {Physical Review Letters}\ }\textbf {\bibinfo {volume} {48}},\ \bibinfo
  {pages} {291--295} (\bibinfo {year} {1982})}\BibitemShut {NoStop}%
\bibitem [{\citenamefont {Navascu\'es}\ \emph {et~al.}(2007)\citenamefont
  {Navascu\'es}, \citenamefont {Pironio},\ and\ \citenamefont
  {Ac\'{\i}n}}]{navascues2007bounding}%
  \BibitemOpen
  \bibfield  {author} {\bibinfo {author} {\bibfnamefont {Miguel}\ \bibnamefont
  {Navascu\'es}}, \bibinfo {author} {\bibfnamefont {Stefano}\ \bibnamefont
  {Pironio}}, \ and\ \bibinfo {author} {\bibfnamefont {Antonio}\ \bibnamefont
  {Ac\'{\i}n}},\ }\bibfield  {title} {\enquote {\bibinfo {title} {Bounding the
  set of quantum correlations},}\ }\href {\doibase
  10.1103/PhysRevLett.98.010401} {\bibfield  {journal} {\bibinfo  {journal}
  {Physical Review Letters}\ }\textbf {\bibinfo {volume} {98}},\ \bibinfo
  {pages} {010401} (\bibinfo {year} {2007})}\BibitemShut {NoStop}%
\bibitem [{\citenamefont {Navascu{\'e}s}\ \emph {et~al.}(2008)\citenamefont
  {Navascu{\'e}s}, \citenamefont {Pironio},\ and\ \citenamefont
  {Ac{\'\i}n}}]{navascues2008convergent}%
  \BibitemOpen
  \bibfield  {author} {\bibinfo {author} {\bibfnamefont {Miguel}\ \bibnamefont
  {Navascu{\'e}s}}, \bibinfo {author} {\bibfnamefont {Stefano}\ \bibnamefont
  {Pironio}}, \ and\ \bibinfo {author} {\bibfnamefont {Antonio}\ \bibnamefont
  {Ac{\'\i}n}},\ }\bibfield  {title} {\enquote {\bibinfo {title} {A convergent
  hierarchy of semidefinite programs characterizing the set of quantum
  correlations},}\ }\href {http://stacks.iop.org/1367-2630/10/i=7/a=073013}
  {\bibfield  {journal} {\bibinfo  {journal} {New Journal of Physics}\ }\textbf
  {\bibinfo {volume} {10}},\ \bibinfo {pages} {073013} (\bibinfo {year}
  {2008})}\BibitemShut {NoStop}%
\bibitem [{\citenamefont {Kogias}\ \emph {et~al.}(2015)\citenamefont {Kogias},
  \citenamefont {Skrzypczyk}, \citenamefont {Cavalcanti}, \citenamefont
  {Ac{\'\i}n},\ and\ \citenamefont {Adesso}}]{kogias2015hierarchy}%
  \BibitemOpen
  \bibfield  {author} {\bibinfo {author} {\bibfnamefont {Ioannis}\ \bibnamefont
  {Kogias}}, \bibinfo {author} {\bibfnamefont {Paul}\ \bibnamefont
  {Skrzypczyk}}, \bibinfo {author} {\bibfnamefont {Daniel}\ \bibnamefont
  {Cavalcanti}}, \bibinfo {author} {\bibfnamefont {Antonio}\ \bibnamefont
  {Ac{\'\i}n}}, \ and\ \bibinfo {author} {\bibfnamefont {Gerardo}\ \bibnamefont
  {Adesso}},\ }\bibfield  {title} {\enquote {\bibinfo {title} {Hierarchy of
  steering criteria based on moments for all bipartite quantum systems},}\
  }\href {\doibase 10.1103/PhysRevLett.115.210401} {\bibfield  {journal}
  {\bibinfo  {journal} {Physical Review Letters}\ }\textbf {\bibinfo {volume}
  {115}},\ \bibinfo {pages} {210401} (\bibinfo {year} {2015})}\BibitemShut
  {NoStop}%
\bibitem [{\citenamefont {Cavalcanti}\ and\ \citenamefont
  {Skrzypczyk}(2017)}]{cavalcanti2016quantum}%
  \BibitemOpen
  \bibfield  {author} {\bibinfo {author} {\bibfnamefont {D}~\bibnamefont
  {Cavalcanti}}\ and\ \bibinfo {author} {\bibfnamefont {P}~\bibnamefont
  {Skrzypczyk}},\ }\bibfield  {title} {\enquote {\bibinfo {title} {Quantum
  steering: a review with focus on semidefinite programming},}\ }\href
  {http://stacks.iop.org/0034-4885/80/i=2/a=024001} {\bibfield  {journal}
  {\bibinfo  {journal} {Reports on Progress in Physics}\ }\textbf {\bibinfo
  {volume} {80}},\ \bibinfo {pages} {024001} (\bibinfo {year}
  {2017})}\BibitemShut {NoStop}%
\bibitem [{\citenamefont {Pironio}\ \emph {et~al.}(2010)\citenamefont
  {Pironio}, \citenamefont {Navascu\'es},\ and\ \citenamefont
  {Ac\'{\i}n}}]{PNA}%
  \BibitemOpen
  \bibfield  {author} {\bibinfo {author} {\bibfnamefont {Stefano}\ \bibnamefont
  {Pironio}}, \bibinfo {author} {\bibfnamefont {Miguel}\ \bibnamefont
  {Navascu\'es}}, \ and\ \bibinfo {author} {\bibfnamefont {Antonio}\
  \bibnamefont {Ac\'{\i}n}},\ }\bibfield  {title} {\enquote {\bibinfo {title}
  {Convergent relaxations of polynomial optimization problems with noncommuting
  variables},}\ }\href {\doibase 10.1137/090760155} {\bibfield  {journal}
  {\bibinfo  {journal} {SIAM J. Optim.}\ }\textbf {\bibinfo {volume} {20}},\
  \bibinfo {pages} {2157--2180} (\bibinfo {year} {2010})}\BibitemShut {NoStop}%
\bibitem [{\citenamefont {Doherty}\ \emph {et~al.}(2008)\citenamefont
  {Doherty}, \citenamefont {Liang}, \citenamefont {Toner},\ and\ \citenamefont
  {Wehner}}]{moments}%
  \BibitemOpen
  \bibfield  {author} {\bibinfo {author} {\bibfnamefont {Andrew~C.}\
  \bibnamefont {Doherty}}, \bibinfo {author} {\bibfnamefont {Yeong-Cherng}\
  \bibnamefont {Liang}}, \bibinfo {author} {\bibfnamefont {Ben}\ \bibnamefont
  {Toner}}, \ and\ \bibinfo {author} {\bibfnamefont {Stephanie}\ \bibnamefont
  {Wehner}},\ }\bibfield  {title} {\enquote {\bibinfo {title} {The quantum
  moment problem and bounds on entangled multi-prover games},}\ }\href
  {\doibase 10.1109/CCC.2008.26} {\bibfield  {journal} {\bibinfo  {journal}
  {Proc. IEEE 23rd Annual Conf. Comp. Compl.}\ ,\ \bibinfo {pages} {199--210}}
  (\bibinfo {year} {2008})}\BibitemShut {NoStop}%
\bibitem [{\citenamefont {Lasserre}(2001)}]{lasserre2001global}%
  \BibitemOpen
  \bibfield  {author} {\bibinfo {author} {\bibfnamefont {Jean~B}\ \bibnamefont
  {Lasserre}},\ }\bibfield  {title} {\enquote {\bibinfo {title} {Global
  optimization with polynomials and the problem of moments},}\ }\href {\doibase
  10.1137/S1052623400366802} {\bibfield  {journal} {\bibinfo  {journal} {SIAM
  Journal on Optimization}\ }\textbf {\bibinfo {volume} {11}},\ \bibinfo
  {pages} {796--817} (\bibinfo {year} {2001})}\BibitemShut {NoStop}%
\bibitem [{\citenamefont {Steeg}\ and\ \citenamefont
  {Galstyan}(2011)}]{steeg2011sequence}%
  \BibitemOpen
  \bibfield  {author} {\bibinfo {author} {\bibfnamefont {Greg~Ver}\
  \bibnamefont {Steeg}}\ and\ \bibinfo {author} {\bibfnamefont {Aram}\
  \bibnamefont {Galstyan}},\ }\bibfield  {title} {\enquote {\bibinfo {title} {A
  sequence of relaxations constraining hidden variable models},}\ }\href
  {https://arxiv.org/abs/1106.1636} {\bibfield  {journal} {\bibinfo  {journal}
  {arXiv:1106.1636}\ } (\bibinfo {year} {2011})}\BibitemShut {NoStop}%
\bibitem [{\citenamefont {Cleve}\ \emph {et~al.}(2004)\citenamefont {Cleve},
  \citenamefont {H{\o}yer}, \citenamefont {Toner},\ and\ \citenamefont
  {Watrous}}]{cleve2004consequences}%
  \BibitemOpen
  \bibfield  {author} {\bibinfo {author} {\bibfnamefont {Richard}\ \bibnamefont
  {Cleve}}, \bibinfo {author} {\bibfnamefont {Peter}\ \bibnamefont {H{\o}yer}},
  \bibinfo {author} {\bibfnamefont {Benjamin}\ \bibnamefont {Toner}}, \ and\
  \bibinfo {author} {\bibfnamefont {John}\ \bibnamefont {Watrous}},\ }\bibfield
   {title} {\enquote {\bibinfo {title} {Consequences and limits of nonlocal
  strategies},}\ }in\ \href {\doibase 10.1109/CCC.2004.1313847} {\emph
  {\bibinfo {booktitle} {Computational Complexity, 2004. Proceedings. 19th IEEE
  Annual Conference on}}}\ (\bibinfo {organization} {IEEE},\ \bibinfo {year}
  {2004})\ pp.\ \bibinfo {pages} {236--249}\BibitemShut {NoStop}%
\bibitem [{Note1()}]{Note1}%
  \BibitemOpen
  \bibinfo {note} {A more formal way to see it is to notice that we are
  restricting the projective algebra of NPA by the commutation condition. Then,
  given that the original algebra already meets the Archimedean condition, the
  convergence holds for the commuting case as well.}\BibitemShut {Stop}%
\bibitem [{\citenamefont {Collins}\ and\ \citenamefont
  {Gisin}(2004)}]{collins2004arelevant}%
  \BibitemOpen
  \bibfield  {author} {\bibinfo {author} {\bibfnamefont {Daniel}\ \bibnamefont
  {Collins}}\ and\ \bibinfo {author} {\bibfnamefont {Nicolas}\ \bibnamefont
  {Gisin}},\ }\bibfield  {title} {\enquote {\bibinfo {title} {A relevant two
  qubit {B}ell inequality inequivalent to the {CHSH} inequality},}\ }\href
  {http://stacks.iop.org/0305-4470/37/i=5/a=021} {\bibfield  {journal}
  {\bibinfo  {journal} {Journal of Physics A: Mathematical and General}\
  }\textbf {\bibinfo {volume} {37}},\ \bibinfo {pages} {1775} (\bibinfo {year}
  {2004})}\BibitemShut {NoStop}%
\bibitem [{\citenamefont {Clauser}\ \emph {et~al.}(1969)\citenamefont
  {Clauser}, \citenamefont {Horne}, \citenamefont {Shimony},\ and\
  \citenamefont {Holt}}]{clauser1969proposed}%
  \BibitemOpen
  \bibfield  {author} {\bibinfo {author} {\bibfnamefont {John~F.}\ \bibnamefont
  {Clauser}}, \bibinfo {author} {\bibfnamefont {Michael~A.}\ \bibnamefont
  {Horne}}, \bibinfo {author} {\bibfnamefont {Abner}\ \bibnamefont {Shimony}},
  \ and\ \bibinfo {author} {\bibfnamefont {Richard~A.}\ \bibnamefont {Holt}},\
  }\bibfield  {title} {\enquote {\bibinfo {title} {Proposed experiment to test
  local hidden-variable theories},}\ }\href {\doibase
  10.1103/PhysRevLett.23.880} {\bibfield  {journal} {\bibinfo  {journal}
  {Physical Review Letters}\ }\textbf {\bibinfo {volume} {23}},\ \bibinfo
  {pages} {880--884} (\bibinfo {year} {1969})}\BibitemShut {NoStop}%
\bibitem [{\citenamefont {Boyd}\ and\ \citenamefont
  {Vandenberghe}(2004)}]{boyd2004convex}%
  \BibitemOpen
  \bibfield  {author} {\bibinfo {author} {\bibfnamefont {Stephen}\ \bibnamefont
  {Boyd}}\ and\ \bibinfo {author} {\bibfnamefont {Lieven}\ \bibnamefont
  {Vandenberghe}},\ }\href
  {https://web.stanford.edu/~boyd/cvxbook/bv_cvxbook.pdf} {\emph {\bibinfo
  {title} {Convex optimization}}}\ (\bibinfo  {publisher} {Cambridge university
  press},\ \bibinfo {year} {2004})\BibitemShut {NoStop}%
\bibitem [{\citenamefont {Gondzio}\ \emph {et~al.}(2014)\citenamefont
  {Gondzio}, \citenamefont {Gruca}, \citenamefont {Hall}, \citenamefont
  {Laskowski},\ and\ \citenamefont {{\.Z}ukowski}}]{gondzio2014solving}%
  \BibitemOpen
  \bibfield  {author} {\bibinfo {author} {\bibfnamefont {Jacek}\ \bibnamefont
  {Gondzio}}, \bibinfo {author} {\bibfnamefont {Jacek~A}\ \bibnamefont
  {Gruca}}, \bibinfo {author} {\bibfnamefont {JA~Julian}\ \bibnamefont {Hall}},
  \bibinfo {author} {\bibfnamefont {Wies{\l}aw}\ \bibnamefont {Laskowski}}, \
  and\ \bibinfo {author} {\bibfnamefont {Marek}\ \bibnamefont {{\.Z}ukowski}},\
  }\bibfield  {title} {\enquote {\bibinfo {title} {Solving large-scale
  optimization problems related to {B}ell's theorem},}\ }\href {\doibase
  http://dx.doi.org/10.1016/j.cam.2013.12.003} {\bibfield  {journal} {\bibinfo
  {journal} {Journal of Computational and Applied Mathematics}\ }\textbf
  {\bibinfo {volume} {263}},\ \bibinfo {pages} {392--404} (\bibinfo {year}
  {2014})}\BibitemShut {NoStop}%
\bibitem [{\citenamefont {Wittek}(2015)}]{algorithm2015wittek}%
  \BibitemOpen
  \bibfield  {author} {\bibinfo {author} {\bibfnamefont {Peter}\ \bibnamefont
  {Wittek}},\ }\bibfield  {title} {\enquote {\bibinfo {title} {Algorithm 950:
  Ncpol2sdpa---sparse semidefinite programming relaxations for polynomial
  optimization problems of noncommuting variables},}\ }\href {\doibase
  10.1145/2699464} {\bibfield  {journal} {\bibinfo  {journal} {ACM Trans. Math.
  Softw.}\ }\textbf {\bibinfo {volume} {41}},\ \bibinfo {pages} {21:1--21:12}
  (\bibinfo {year} {2015})}\BibitemShut {NoStop}%
\bibitem [{Note2()}]{Note2}%
  \BibitemOpen
  \bibinfo {note} {Available at \protect \url
  {http://www.mosek.com/}}\BibitemShut {NoStop}%
\bibitem [{Note3()}]{Note3}%
  \BibitemOpen
  \bibinfo {note} {The code used is available under an open source license at
  \protect \url
  {https://github.com/FlavioBaccari/Hierarchy-for-nonlocality-detection}}\BibitemShut
  {NoStop}%
\bibitem [{\citenamefont {Wallman}\ \emph {et~al.}(2011)\citenamefont
  {Wallman}, \citenamefont {Liang},\ and\ \citenamefont
  {Bartlett}}]{wallman2011generating}%
  \BibitemOpen
  \bibfield  {author} {\bibinfo {author} {\bibfnamefont {Joel~J}\ \bibnamefont
  {Wallman}}, \bibinfo {author} {\bibfnamefont {Yeong-Cherng}\ \bibnamefont
  {Liang}}, \ and\ \bibinfo {author} {\bibfnamefont {Stephen~D}\ \bibnamefont
  {Bartlett}},\ }\bibfield  {title} {\enquote {\bibinfo {title} {Generating
  nonclassical correlations without fully aligning measurements},}\ }\href
  {\doibase 10.1103/PhysRevA.83.022110} {\bibfield  {journal} {\bibinfo
  {journal} {Physical Review A}\ }\textbf {\bibinfo {volume} {83}},\ \bibinfo
  {pages} {022110} (\bibinfo {year} {2011})}\BibitemShut {NoStop}%
\bibitem [{\citenamefont {McKague}(2011)}]{mckague2011self}%
  \BibitemOpen
  \bibfield  {author} {\bibinfo {author} {\bibfnamefont {Matthew}\ \bibnamefont
  {McKague}},\ }\bibfield  {title} {\enquote {\bibinfo {title} {Self-testing
  graph states},}\ }in\ \href {\doibase 10.1007/978-3-642-54429-3_7} {\emph
  {\bibinfo {booktitle} {Conference on Quantum Computation, Communication, and
  Cryptography}}}\ (\bibinfo {organization} {Springer},\ \bibinfo {year}
  {2011})\ pp.\ \bibinfo {pages} {104--120}\BibitemShut {NoStop}%
\bibitem [{\citenamefont {{Hein}}\ \emph {et~al.}(2006)\citenamefont {{Hein}},
  \citenamefont {{D{\"u}r}}, \citenamefont {{Eisert}}, \citenamefont
  {{Raussendorf}}, \citenamefont {{Van den Nest}},\ and\ \citenamefont
  {{Briegel}}}]{hein2006entanglement}%
  \BibitemOpen
  \bibfield  {author} {\bibinfo {author} {\bibfnamefont {M.}~\bibnamefont
  {{Hein}}}, \bibinfo {author} {\bibfnamefont {W.}~\bibnamefont {{D{\"u}r}}},
  \bibinfo {author} {\bibfnamefont {J.}~\bibnamefont {{Eisert}}}, \bibinfo
  {author} {\bibfnamefont {R.}~\bibnamefont {{Raussendorf}}}, \bibinfo {author}
  {\bibfnamefont {M.}~\bibnamefont {{Van den Nest}}}, \ and\ \bibinfo {author}
  {\bibfnamefont {H.~.}\ \bibnamefont {{Briegel}}},\ }\bibfield  {title}
  {\enquote {\bibinfo {title} {{Entanglement in Graph States and its
  Applications}},}\ }\href {https://arxiv.org/abs/quant-ph/0602096} {\bibfield
  {journal} {\bibinfo  {journal} {arXiv:quant-ph/0602096}\ } (\bibinfo {year}
  {2006})}\BibitemShut {NoStop}%
\bibitem [{\citenamefont {Lanyon}\ \emph {et~al.}(2014)\citenamefont {Lanyon},
  \citenamefont {Zwerger}, \citenamefont {Jurcevic}, \citenamefont {Hempel},
  \citenamefont {D{\"u}r}, \citenamefont {Briegel}, \citenamefont {Blatt},\
  and\ \citenamefont {Roos}}]{lanyon2014experimental}%
  \BibitemOpen
  \bibfield  {author} {\bibinfo {author} {\bibfnamefont {B.~P.}\ \bibnamefont
  {Lanyon}}, \bibinfo {author} {\bibfnamefont {M.}~\bibnamefont {Zwerger}},
  \bibinfo {author} {\bibfnamefont {P.}~\bibnamefont {Jurcevic}}, \bibinfo
  {author} {\bibfnamefont {C.}~\bibnamefont {Hempel}}, \bibinfo {author}
  {\bibfnamefont {W.}~\bibnamefont {D{\"u}r}}, \bibinfo {author} {\bibfnamefont
  {H.~J.}\ \bibnamefont {Briegel}}, \bibinfo {author} {\bibfnamefont
  {R.}~\bibnamefont {Blatt}}, \ and\ \bibinfo {author} {\bibfnamefont {C.~F.}\
  \bibnamefont {Roos}},\ }\bibfield  {title} {\enquote {\bibinfo {title}
  {Experimental violation of multipartite {B}ell inequalities with trapped
  ions},}\ }\href {\doibase 10.1103/PhysRevLett.112.100403} {\bibfield
  {journal} {\bibinfo  {journal} {Physical Review Letters}\ }\textbf {\bibinfo
  {volume} {112}},\ \bibinfo {pages} {100403} (\bibinfo {year}
  {2014})}\BibitemShut {NoStop}%
\bibitem [{\citenamefont {Monz}\ \emph {et~al.}(2011)\citenamefont {Monz},
  \citenamefont {Schindler}, \citenamefont {Barreiro}, \citenamefont {Chwalla},
  \citenamefont {Nigg}, \citenamefont {Coish}, \citenamefont {Harlander},
  \citenamefont {H\"ansel}, \citenamefont {Hennrich},\ and\ \citenamefont
  {Blatt}}]{monz201114qubit}%
  \BibitemOpen
  \bibfield  {author} {\bibinfo {author} {\bibfnamefont {Thomas}\ \bibnamefont
  {Monz}}, \bibinfo {author} {\bibfnamefont {Philipp}\ \bibnamefont
  {Schindler}}, \bibinfo {author} {\bibfnamefont {Julio~T.}\ \bibnamefont
  {Barreiro}}, \bibinfo {author} {\bibfnamefont {Michael}\ \bibnamefont
  {Chwalla}}, \bibinfo {author} {\bibfnamefont {Daniel}\ \bibnamefont {Nigg}},
  \bibinfo {author} {\bibfnamefont {William~A.}\ \bibnamefont {Coish}},
  \bibinfo {author} {\bibfnamefont {Maximilian}\ \bibnamefont {Harlander}},
  \bibinfo {author} {\bibfnamefont {Wolfgang}\ \bibnamefont {H\"ansel}},
  \bibinfo {author} {\bibfnamefont {Markus}\ \bibnamefont {Hennrich}}, \ and\
  \bibinfo {author} {\bibfnamefont {Rainer}\ \bibnamefont {Blatt}},\ }\bibfield
   {title} {\enquote {\bibinfo {title} {14-qubit entanglement: Creation and
  coherence},}\ }\href {\doibase 10.1103/PhysRevLett.106.130506} {\bibfield
  {journal} {\bibinfo  {journal} {Physical Review Letters}\ }\textbf {\bibinfo
  {volume} {106}},\ \bibinfo {pages} {130506} (\bibinfo {year}
  {2011})}\BibitemShut {NoStop}%
\bibitem [{\citenamefont {Wang}\ \emph {et~al.}(2016)\citenamefont {Wang},
  \citenamefont {Chen}, \citenamefont {Li}, \citenamefont {Huang},
  \citenamefont {Liu}, \citenamefont {Chen}, \citenamefont {Luo}, \citenamefont
  {Su}, \citenamefont {Wu}, \citenamefont {Li}, \citenamefont {Lu},
  \citenamefont {Hu}, \citenamefont {Jiang}, \citenamefont {Peng},
  \citenamefont {Li}, \citenamefont {Liu}, \citenamefont {Chen}, \citenamefont
  {Lu},\ and\ \citenamefont {Pan}}]{wang2016experimental}%
  \BibitemOpen
  \bibfield  {author} {\bibinfo {author} {\bibfnamefont {Xi-Lin}\ \bibnamefont
  {Wang}}, \bibinfo {author} {\bibfnamefont {Luo-Kan}\ \bibnamefont {Chen}},
  \bibinfo {author} {\bibfnamefont {W.}~\bibnamefont {Li}}, \bibinfo {author}
  {\bibfnamefont {H.-L.}\ \bibnamefont {Huang}}, \bibinfo {author}
  {\bibfnamefont {C.}~\bibnamefont {Liu}}, \bibinfo {author} {\bibfnamefont
  {C.}~\bibnamefont {Chen}}, \bibinfo {author} {\bibfnamefont {Y.-H.}\
  \bibnamefont {Luo}}, \bibinfo {author} {\bibfnamefont {Z.-E.}\ \bibnamefont
  {Su}}, \bibinfo {author} {\bibfnamefont {D.}~\bibnamefont {Wu}}, \bibinfo
  {author} {\bibfnamefont {Z.-D.}\ \bibnamefont {Li}}, \bibinfo {author}
  {\bibfnamefont {H.}~\bibnamefont {Lu}}, \bibinfo {author} {\bibfnamefont
  {Y.}~\bibnamefont {Hu}}, \bibinfo {author} {\bibfnamefont {X.}~\bibnamefont
  {Jiang}}, \bibinfo {author} {\bibfnamefont {C.-Z.}\ \bibnamefont {Peng}},
  \bibinfo {author} {\bibfnamefont {L.}~\bibnamefont {Li}}, \bibinfo {author}
  {\bibfnamefont {N.-L.}\ \bibnamefont {Liu}}, \bibinfo {author} {\bibfnamefont
  {Yu-Ao}\ \bibnamefont {Chen}}, \bibinfo {author} {\bibfnamefont {Chao-Yang}\
  \bibnamefont {Lu}}, \ and\ \bibinfo {author} {\bibfnamefont {Jian-Wei}\
  \bibnamefont {Pan}},\ }\bibfield  {title} {\enquote {\bibinfo {title}
  {Experimental ten-photon entanglement},}\ }\href {\doibase
  10.1103/PhysRevLett.117.210502} {\bibfield  {journal} {\bibinfo  {journal}
  {Physical Review Letters}\ }\textbf {\bibinfo {volume} {117}},\ \bibinfo
  {pages} {210502} (\bibinfo {year} {2016})}\BibitemShut {NoStop}%
\bibitem [{\citenamefont {Chen}\ \emph {et~al.}(2017)\citenamefont {Chen},
  \citenamefont {Li}, \citenamefont {Yao}, \citenamefont {Huang}, \citenamefont
  {Li}, \citenamefont {Lu}, \citenamefont {Yuan}, \citenamefont {Zhang},
  \citenamefont {Jiang}, \citenamefont {Peng}, \citenamefont {Li},
  \citenamefont {Liu}, \citenamefont {Ma}, \citenamefont {Lu}, \citenamefont
  {Chen},\ and\ \citenamefont {Pan}}]{chen2017observation}%
  \BibitemOpen
  \bibfield  {author} {\bibinfo {author} {\bibfnamefont {Luo-Kan}\ \bibnamefont
  {Chen}}, \bibinfo {author} {\bibfnamefont {Zheng-Da}\ \bibnamefont {Li}},
  \bibinfo {author} {\bibfnamefont {Xing-Can}\ \bibnamefont {Yao}}, \bibinfo
  {author} {\bibfnamefont {Miao}\ \bibnamefont {Huang}}, \bibinfo {author}
  {\bibfnamefont {Wei}\ \bibnamefont {Li}}, \bibinfo {author} {\bibfnamefont
  {He}~\bibnamefont {Lu}}, \bibinfo {author} {\bibfnamefont {Xiao}\
  \bibnamefont {Yuan}}, \bibinfo {author} {\bibfnamefont {Yan-Bao}\
  \bibnamefont {Zhang}}, \bibinfo {author} {\bibfnamefont {Xiao}\ \bibnamefont
  {Jiang}}, \bibinfo {author} {\bibfnamefont {Cheng-Zhi}\ \bibnamefont {Peng}},
  \bibinfo {author} {\bibfnamefont {Li}~\bibnamefont {Li}}, \bibinfo {author}
  {\bibfnamefont {Nai-Le}\ \bibnamefont {Liu}}, \bibinfo {author}
  {\bibfnamefont {Xiongfeng}\ \bibnamefont {Ma}}, \bibinfo {author}
  {\bibfnamefont {Chao-Yang}\ \bibnamefont {Lu}}, \bibinfo {author}
  {\bibfnamefont {Yu-Ao}\ \bibnamefont {Chen}}, \ and\ \bibinfo {author}
  {\bibfnamefont {Jian-Wei}\ \bibnamefont {Pan}},\ }\bibfield  {title}
  {\enquote {\bibinfo {title} {Observation of ten-photon entanglement using
  thin {BiB3O6} crystals},}\ }\href {\doibase 10.1364/OPTICA.4.000077}
  {\bibfield  {journal} {\bibinfo  {journal} {Optica}\ }\textbf {\bibinfo
  {volume} {4}},\ \bibinfo {pages} {77--83} (\bibinfo {year}
  {2017})}\BibitemShut {NoStop}%
\end{thebibliography}%

\appendix
\section{Details of the method}\label{app:details}

Here, we present in more detail the SDP relaxation associated to quantum realizations with commuting measurements. In order to be consistent with the examples presented in the main text, we express it in terms of correlators, but we stress that a formulation in terms of projector and probabilities for higher numbers of outcomes is straightforward.

Let us consider that the $N$ observers $A_i$ are allowed to perform $m$ dicothomic measurements each. We can therefore define the operators $M^{(i)}_{x_i} = M^{1}_{x_i} - M^{0}_{x_i}$ in terms of measurements $M^{a_i}_{x_i}$. It can be easily seen that expectation values of $ M^{(i)}_{x_i}$ correspond to the correlators (\ref{corr}).

For any quantum realization of such operators, it is possible to show that they satisfy the following properties:

\begin{enumerate}[i)]

\item $(M^{(i)}_{x_i})^{\dagger}$ = $M^{(i)}_{x_i}$ for any $i = 1,{\ldots}N$ and $x_i = 1,{\ldots},m$,

\item  $(M^{(i)}_{x_i})^2 = \mathbb{1}$ for any $i = 1,{\ldots}N$ and $x_i = 1,{\ldots},m$,

\item $[M^{(i)}_{x_i},  M^{(j)}_{x_j} ] = 0 \, \, \, $ for any $i \neq j$ and $x_i,x_j = 1,{\ldots},m$.
\end{enumerate}

Now, let us consider that the sets $\mathcal{O}_{\nu}$ we introduce in section~\ref{sec:method} consist exactly of all the products of the $\lbrace M^{(i)}_{x_i} \rbrace$ up to order $\nu$. Then, by indexing the operators in the sets as $\mathcal{O}_i$ for $i = 1,{\ldots},k$, we define the $k \times k$ moment matrix as follows

\begin{equation*}
\Gamma_{ij} = tr (\rho_N \mathcal{O}^{\dagger}_i \mathcal{O}_j )
\end{equation*}

where $\rho_N$ is a generic N-partite quantum state. As it was shown in~\citep{navascues2007bounding,navascues2008convergent}, for any set of quantum correlations $P$, the properties i)-iii) and the fact that the associated $\rho_N$ is a proper quantum state reflect into the following properties of the moment matrix:

\begin{itemize}
\item $\Gamma^{\dagger} = \Gamma$,
\item $\Gamma \succeq 0$,
\item the entries of the matrix are constrained by some linear equations of the form
\begin{equation*}
\sum_{i,j} (F_m)_{ij} \Gamma_{ij} = g_m(P) \, \, \, \, m = 1,{\ldots},l
\end{equation*}
where $(F_m)_{ij}$ are some coefficients and the $g_m(P)$ can depend on the values of the correlators composing the $P$ vector, as such

 \begin{multline*}
g_m(P) = (g_m)_0 +
\sum_{k = 1}^{N} \sum_{\substack{i_1 < {\ldots} < i_k \\ i_1,{\ldots},i_j}} (g_m)^{i_1,{\ldots},i_k}_{j_1,{\ldots}j_k} \langle M_{j_1}^{(i_1)}{\ldots} M_{j_k}^{(i_k)}   \rangle
\end{multline*}

\end{itemize}

Up to this point, the method we describe coincides with the NPA hierarchy~\citep{navascues2007bounding,navascues2008convergent}, which is used to check whether a set of observed correlations is compatible with a quantum realization. In order to define a hierarchy to test for local hidden variables realization, we introduce the additional condition that all the measurements for the same party have to also be commuting, namely

\begin{enumerate}[i)]
\setcounter{enumi}{3}
\item $[ M_{x_i}^{(i)} , M_{y_i}^{(i)} ] = 0 \, \, \ $ for any $i = 1,{\ldots},N$ and $x_i \neq y_i = 1,{\ldots},m$.
\end{enumerate}

It can be seen that property iv) implies a second set of linear constraints on the $\Gamma$ matrix, which we identify as

\begin{equation*}
\sum_{i,j} (F{'}_m)_{ij} \Gamma_{ij} = g{'}_m(P) \, \, \, \, m = 1,{\ldots},l{'}
\end{equation*}
To make it clearer, we show an example of linear constraint that can come only if we impose condition iv). Let us consider the following four operators: $\mathcal{O}_k =  M_{x_i}^{(i)}   M_{y_i}^{(i)}$, $\mathcal{O}_l =  M_{x_i}^{(i)}   M_{x_j}^{(j)}$,
$\mathcal{O}_n =  M_{y_i}^{(i)} $ and $\mathcal{O}_m =  M_{x_j}^{(j)} $. It is easy to see that, by exploiting i)-iii) plus iv), $\Gamma_{kl} = \Gamma_{nm}$ for any choice of $x_i,y_i,x_j = 1,{\ldots},m$ and $i,j = 1,{\ldots},N$.

Now, for any chosen $\mathcal{O}_{\nu}$, we can test whether an observed distribution $P$ is compatible with a local model via the following SDP

\begin{align}
 \text{maximize} \, {} & \, \, \,  \lambda , \nonumber \\
       \text{subject to} {} & \,\,\, \Gamma - \lambda \mathbb{1} \succeq 0 , \nonumber \\
       & \, \,  \sum_{i,j} (F_m)_{ij} \Gamma_{ij} = g_m(P) \, \, \, \, m = 1,{\ldots},l  \, \, ,\\
     & \, \, \sum_{i,j} (F{'}_m)_{ij} \Gamma_{ij} = g{'}_m(P) \, \, \, \, m = 1,{\ldots},l{'} \, \, , \nonumber
\label{primal}
\end{align}
which is the \textit{primal form} of the problem. A solution $\lambda^{\ast}_{min} < 0 $ implies that it is not possible to find a semidefinite positive moment matrix satisfying the given linear constraints. Therefore $P$ has no quantum realization with commuting measurements and we conclude it is nonlocal. We notice that by increasing the value of $\nu$ we get a sequence of more and more stringent tests for nonlocality. Indeed, the linear constraints for the level $\nu$ are always a subset of the ones coming from $\nu +1$.
Moreover, in analogy with the NPA hierarchy, the series of tests is convergent; hence any nonlocal correlation will give a negative solution $\lambda^{\ast}_{min}$ at a finite step of the sequence.

Interestingly, we can also study the \textit{dual form} of the SDP problem, which reads as follows:

\begin{align}
 \text{minimize} \, {} & \, \, \,  G(P) = \sum_k y_k g_k(P) + \sum_k y{'}_k g{'}_k(P) , \nonumber \\
       \text{subject to} {} & \,\,\,  \sum_k y_k F^{T}_k + \sum_k y{'}_k F{'}^{T}_k \succeq 0 , \\
     & \, \, \,  \sum_k y_k \text{tr}(F^{T}_k) + \sum_k y{'}_k \text{tr}(F{'}^{T}_k) = 1. \nonumber
\end{align}
Thanks to the strong duality of the problem, a negative solution for the primal implies also $G(P) = \lambda^{\ast}_{min} < 0$. Since any point in $\mathcal{L}_{\nu}$ satisfies the SDP condition at level $\nu$ with $G(P) \geq 0$, we can interpret $G(P)$ as a Bell-like inequality separating the $\mathcal{L}_{\nu}$ from the rest of the correlations. Indeed, since $g_k (P)$ and  $g{'}_k (P)$ are linear in terms of the probability distribution, we derive that $G(P) \geq 0$ defines also a linear inequality for $P$. Violation of such an inequality directly implies nonlocality.

\section{Proof of local bound and quantum violation for the inequalities}\label{app:ineq}

We start by proving the local bounds for the inequalities introduced in the main text. To do so, we remind the reader that to derive the maximal value attained by local correlations it is enough to maximize over the vertices of the local set. In the correlator space, the deterministic local strategies (DLS) take the form

\begin{equation}
\langle M_{j_1}^{(i_1)}{\ldots} M_{j_k}^{(i_k)}  \rangle = \langle M_{j_1}^{(i_1)} \rangle {\ldots} \langle M_{j_k}^{(i_k)}  \rangle
\end{equation}
where each $ M_{x_i}^{(i)} $ term can take only $1$ and $-1$ values.
By using this property, inequality (\ref{ineq3}) becomes

\begin{equation}
\begin{split}
\mathcal{I}_{mix}^3(DLS) =  (N-1) \left[ \langle M_{1}^{(1)} \rangle  + \langle M_{0}^{(1)} \rangle \right] \mathcal{T}_0 \\ + \left[ \langle M_{1}^{(1)} \rangle  - \langle M_{0}^{(1)} \rangle \right] \mathcal{T}_2
\end{split}
\end{equation}
where $\mathcal{T}_0 =  \langle M_{0}^{(2)} \rangle {\ldots} \langle M_{0}^{(N)}  \rangle $ and $\mathcal{T}_2 = \sum_{i = 2}^{N} \langle M_{2}^{(i)} \rangle$. For any number of parties $N$, we have that $\mathcal{T}_0 \leq 1$ and $\mathcal{T}_2 \leq N-1$; therefore

\begin{equation}
\mathcal{I}_{mix}^3(DLS) \leq 2 (N-1)  \langle M_{1}^{(1)} \rangle \leq 2(N-1)
\end{equation}
Similarly, for any deterministic strategy, inequality (\ref{ineq2}) takes the from

\begin{equation}
\begin{split}
\mathcal{I}_{mix}^2(DLS) =  (N-1) \left[ \langle M_{1}^{(1)} \rangle  + \langle M_{0}^{(1)} \rangle \right] \mathcal{T}_0 \\ + \left[ \langle M_{1}^{(1)} \rangle  - \langle M_{0}^{(1)} \rangle \right] \mathcal{T}_1
\end{split}
\end{equation}
where $\mathcal{T}_1 = \sum_{i = 2}^{N} \langle M_{1}^{(i)} \rangle$. As for before, we can use the argument that $\mathcal{T}_0 \leq 1$ and $\mathcal{T}_1 \leq N-1$
to conclude

\begin{equation}
\mathcal{I}_{mix}^2(DLS) \leq 2 (N-1)  \langle M_{1}^{(1)} \rangle \leq 2(N-1)
\end{equation}

Regarding the quantum violation, we recall that the scenario we consider is $| \psi \rangle = | GHZ_N \rangle = \frac{1}{\sqrt{2}} (|0 \rangle^{\otimes N} + |1 \rangle^{\otimes N} )$ with measurements choices $M_0^{(i)} = \sigma_x$, $M_1^{(i)} = \sigma_d = \frac{1}{\sqrt{2}} (\sigma_x + \sigma_z)$ and $M_2^{(i)} = \sigma_z$ for all $i = 1,{\ldots},N$.
It is easy to check that, for a $GHZ$ state of any number of parties, the following is true

\begin{itemize}
\item $\langle \sigma_{x}^{(i)}  \sigma_{x}^{(j)} \rangle =  \langle \sigma_{x}^{(i)}  \sigma_{z}^{(j)} \rangle = 0$ for any $i \neq j = 1,{\ldots},N$.
\item $\langle \sigma_{z}^{(i)}  \sigma_{z}^{(j)} \rangle = 1$ and therefore $\langle \sigma_{d}^{(i)}  \sigma_{z}^{(j)} \rangle = \frac{1}{\sqrt{2}}$ and $\langle \sigma_{d}^{(i)}  \sigma_{d}^{(j)} \rangle = \frac{1}{2}$for any $i \neq j = 1,{\ldots},N$.
\item $\langle \sigma_{x}^{(1)}  \sigma_{x}^{(2)} {\ldots}   \sigma_{x}^{(N)} \rangle = 1$ and $\langle \sigma_{z}^{(1)}  \sigma_{x}^{(2)} {\ldots}   \sigma_{x}^{(N)} \rangle = 0$, therefore  $\langle \sigma_{d}^{(1)}  \sigma_{x}^{(2)} {\ldots}   \sigma_{x}^{(N)} \rangle = \frac{1}{\sqrt{2}}$ for any $N$.

\end{itemize}
By using the properties listed above, one can check that

\begin{equation}
\langle \mathcal{I}_{mix}^{3} \rangle_{GHZ_{N}} = (1+ \sqrt{2}) (N -1) \approx 2.41(N-1)
\end{equation}
and, similarly, that

\begin{equation}
\langle \mathcal{I}_{mix}^{2} \rangle_{GHZ_{N}} = \frac{3+ \sqrt{2}}{2} (N -1) \approx 2.21(N-1)
\end{equation}
Moreover, we notice that by changing the measurement setting, one can achieve a higher violation of $\mathcal{I}_{mix}^2$. Indeed, it is easy to see that by choosing $M_0^{(1)} = \frac{1}{\sqrt{2}} (\sigma_x + \sigma_z)$,  $M_1^{(1)} = \frac{1}{\sqrt{2}} (\sigma_x - \sigma_z)$ and $M_0^{(i)} = \sigma_x$,$M_1^{(i)} = \sigma_z$ for $i = 2,{\ldots},N$, the resulting violation is

\begin{equation}
\langle \mathcal{I}_{mix}^{2} \rangle_{GHZ_{N}} = 2 \sqrt{2} (N -1) \approx 2.83(N-1)
\end{equation}

To conclude, we proceed with the analysis of the robustness to noise. We recall that this implies considering the noisy version of the $GHZ$ state; namely

\begin{equation}
\rho_N (p) = (1 - p) \rho_{GHZ_N} + p \frac{\mathbb{1}_N}{2^N}
\end{equation}
where $0 \leq p \leq 1$ represent the amount of white noise added to the state. It can be easily seen that the noise affects the values of the correlators for the $GHZ$ state in the following way

\begin{equation}
\langle \sigma_{j_1}^{(i_1)}{\ldots} \sigma_{j_k}^{(i_k)} \rangle_{\rho_N} = (1-p) \langle \sigma_{j_1}^{(i_1)}{\ldots} \sigma_{j_k}^{(i_k)} \rangle_{GHZ_N}
\end{equation}
for any $j_l \in \lbrace x,y,z \rbrace$ and $1 \leq k \leq N$. Therefore we can consider the noise as a simple damping factor in the violation of the inequalities. By using this fact, we get that $\mathcal{I}_{mix}^{3}$ is violated as long as $(1-p)(1 + \sqrt{2}) (N-1) > 2 (N-1)$; hence

\begin{equation}
p_{max}( \mathcal{I}_{mix}^{3} ) = \frac{\sqrt{2} -1 }{\sqrt{2} + 1} \approx 0.17
\end{equation}

By the same argument, we analyze $\mathcal{I}_{mix}^{2}$ for the two measurements settings that we introduce. For the first one, the inequality is violated as long as $(1-p)\frac{3 + \sqrt{2}}{2} (N-1) > 2 (N-1)$ and, therefore,

\begin{equation}
p_{max}( \mathcal{I}_{mix}^{2} ) = \frac{\sqrt{2} -1 }{\sqrt{2} + 3} \approx 0.09
\end{equation}
while for the second one, the violation is preserved for $(1-p)2 \sqrt{2}(N-1) > 2 (N-1)$; hence

\begin{equation}
p'_{max}( \mathcal{I}_{mix}^{2} ) = 1 - \frac{\sqrt{2}}{2} \approx 0.29
\end{equation}
Clearly, we see that for the second setting a higher violation results also in a significantly higher robustness to noise.

\end{document}